\documentclass[
reprint,
amsmath,amssymb,
prb,
]{revtex4-2}

\usepackage{graphicx}
\usepackage{dcolumn}
\usepackage{bm}
\usepackage{hyperref}

\begin{document}

\title{Melting phase relation of seifertite and pyrite-type SiO\textsubscript{2}\\ determined by machine learning potentials}%

\author{Doyoon Park$^{1}$}
\author{Xin Deng$^{2}$}
\email{xdeng@carnegiescience.edu}
\author{Jie Deng$^{3}$}
\email{jie.deng@princeton.edu}
\affiliation{
\textsuperscript{1}\mbox{Department of Mechanical and Aerospace Engineering, Princeton University, Princeton, New Jersey 08544, USA} \\
\textsuperscript{2}\mbox{Earth and Planets Laboratory, Carnegie Institution for Science, Washington, DC, 20015, USA} \\
\textsuperscript{3}Department of Geosciences, Princeton University, Princeton, New Jersey 08544, USA
}

\date{\today}

\begin{abstract}
Silica (SiO$_2$) is fundamental to both industrial technology and planetary science, yet the phase relations of its high-pressure polymorphs remain poorly constrained. Here, we develop two machine learning potentials (MLPs) for SiO$_2$ that faithfully represent the SCAN and PBEsol exchange-correlation functionals over a wide temperature (1000–10000 K) and pressure (100–400 GPa) range using deep neural networks. With large-scale two-phase simulations powered by these potentials, we determine the melting curves of seifertite and pyrite-type SiO$_2$ and infer the solid-solid phase boundary between these two phases. The SCAN functional, which captures intermediate-range van der Waals interactions, reproduces structural and thermodynamic properties with high fidelity, predicting melting temperatures 6–10 \% higher and a seifertite to pyrite-type transition pressure 22 \% higher than the PBEsol. The strongly negative Clapeyron slope (–6.1 MPa/K) of this transition suggests that mantle convection could be highly layered in super-Earth exoplanets, potentially affecting their long-term thermal evolution and habitability.
\end{abstract}

\maketitle

\section{\label{sec:level1}Introduction}
Silica (SiO$_2$) is one of the most abundant materials in the crust and mantle of rocky planets and has wide industrial applications ranging from toothpaste to semiconductors, making it a key substance in solid-state physics, geosciences, and engineering. Despite its simple chemical formula, SiO$_2$ exhibits extensive polymorphism and a rich high-pressure phase diagram \cite{kuwayama2005, oganov2005, liu2021}. Unraveling the phase relations of these high-pressure polymorphs is crucial for understanding the formation and long-term evolution of rocky exoplanetary interiors. This study focuses on the seifertite ($\alpha$-PbO$_2$-type, $Pbcn$) and pyrite-type ($Pa\bar{3}$) SiO$_2$, which are poorly studied but important for understanding Earth's and exoplanetary mantles.

Despite extensive experimental efforts, the melting behaviors of these two SiO$_2$ phases at high pressures remain uncertain. For example, laser-heated diamond anvil cell (DAC) experiments report a melting temperature of $\sim$6200 K at $\sim$120 GPa \cite{andrault2020, andrault2022}, which is approximately 1100 K higher than values inferred from shock-compression studies \cite{millot2015}. In addition to the melting behavior, the solid-solid phase boundaries between high-pressure polymorphs are also unclear, especially at high temperatures \cite{kuwayama2005}. 

To complement experiments and overcome their limitations under extreme conditions, computational approaches based on density functional theory (DFT) \cite{kohn1965} have been widely applied. A recent study extended the SiO$_2$ melting curve to pressures up to 500 GPa \cite{geng2024}, finding no abrupt change in slope as reported in previous theoretical studies \cite{gonzalez2016, gonzalez2016conf}. In addition, none of these studies examines the solid-solid phase boundaries. Previous theoretical estimates of the seifertite to pyrite-type transition pressure range from 201 GPa \cite{oganov2005} to 215 GPa \cite{das2020} at 0 K, much lower than the experimental value of $\sim$260 GPa \cite{kuwayama2005}. Notably, the precise location of the triple point, where the melting curve intersects the solid-solid phase boundary, remains undetermined.

Furthermore, previous theoretical studies on the SiO$_2$ melting curve are based on local density approximation (LDA) \cite{ceperley1980, perdew1981} or generalized gradient approximation (GGA) \cite{lee1988, becke1988, perdew1996} exchange-correlation (XC) functionals. While these approaches are computationally efficient, they are generally less accurate than meta-GGA and hybrid functionals according to Jacob’s ladder of DFT \cite{perdew2001}. Recent studies have demonstrated that the strongly constrained and appropriately normed (SCAN) meta-GGA functional \cite{sun2015} provides more accurate predictions of structural and thermodynamic properties, including lattice parameters \cite{sun2015}, formation energies \cite{sun2016}, lattice dynamics \cite{ning2022}, and melting temperatures \cite{rang2019,jinnouchi2019}, than GGA functionals such as Perdew-Burke-Ernzerhof (PBE) \cite{perdew1996}.

Although SCAN offers high accuracy, its computational cost makes direct \textit{ab initio} molecular dynamics (AIMD) simulations of large systems with thousands of atoms impractical. Machine learning potentials (MLPs) offer \textit{ab initio}-level accuracy at a fraction of the cost, enabling simulations of large systems over long timescales \cite{behler2007, zhang2018}. Once training and testing datasets are generated to build an MLP, no subsequent DFT calculations are required, allowing computationally expensive functionals such as SCAN to be effectively approximated. This efficiency makes MLPs ideally suited for two-phase simulations of large systems containing thousands of atoms over extended timescales on the order of nanoseconds, substantially reducing uncertainty in the predicted melting temperatures. Developing a reliable MLP that spans multiple phases across a wide range of pressure and temperature remains challenging \cite{yang2021}, but can be achieved through enhanced sampling with a well-designed set of collective variables (CVs) \cite{deng2023}. 

In this study, we develop two robust and transferable MLPs for SiO$_2$, each trained on data derived from the SCAN and PBEsol XC functionals \cite{perdew2008}, respectively. These potentials accurately capture interatomic interactions over a wide pressure range (100–400 GPa) and temperature range (1000–10000 K). Using large-scale two-phase coexistence simulations powered by these MLPs, we determine the melting curves of seifertite and pyrite-type SiO$_2$ and locate their triple point. We then use this triple point to constrain the solid-solid phase boundary between the two polymorphs.

\section{Method}
A machine learning potential is a neural network model that approximates the interatomic potential-energy surface. Our workflow for determining the melting of seifertite and pyrite-type SiO$_2$ involves three main steps. First, we generate an initial training set using configurations sampled from an existing SiO$_2$ potential valid at lower pressures and temperatures, and compute their energies and forces with \textit{ab initio} calculations to train a preliminary MLP. Second, we perform MD simulations with this intermediate MLP, applying enhanced sampling in the target pressure and temperature range, to extract new atomic configurations that are added to the training data, and retrain the model using the updated dataset. This step is iteratively repeated to obtain a robust MLP, following the procedure described in Deng \textit{et al.} \cite{deng2023}. Lastly, we conduct two-phase coexistence simulations with the MLP to capture the melting transitions of the seifertite and pyrite structures. Detailed procedures are described in the following sections.

\subsection{Enhanced sampling}
Enhanced sampling accelerates the exploration of complex free-energy surfaces (FES), including those with multiple phases, interfaces, and rare transition states, by introducing a bias potential along well-defined collective variables (CVs). The multithermal-multibaric (MTMB) simulation is an enhanced sampling technique that enables uniform sampling in both energy and volume over specified temperature and pressure ranges. It builds upon the variationally enhanced sampling (VES) framework \cite{valsson2014}, which defines a functional of the bias potential $V(\mathbf{s})$ as
\begin{equation}
\Omega[V(\mathbf{s})] = \frac{1}{\beta} \ln 
\frac{\int d\mathbf{s} \, e^{-\beta \left[F(\mathbf{s}) + V(\mathbf{s})\right]}}
     {\int d\mathbf{s} \, e^{-\beta F(\mathbf{s})}} 
+ \int d\mathbf{s} \, p(\mathbf{s})V(\mathbf{s}),
\end{equation}
where $\mathbf{s}\left(\mathbf{R}\right)$ is the set of collective variables as a function of the atomic coordinates $\mathbf{R}$, $\beta=\left(k_BT\right)^{-1}$ is the inverse temperature, with the Boltzmann constant $k_B$ and temperature $T$, and $p\left(\mathbf{s}\right)$ is an arbitrary probability distribution. The Helmholtz free energy $F\left(\mathbf{s}\right)$ is given as 
\begin{equation}
F(\mathbf{s}) = -\frac{1}{\beta} \ln \int d\mathbf{R} \, \delta \big(\mathbf{s} - \mathbf{s}(\mathbf{R})\big) \,
e^{-\beta U(\mathbf{R})},
\end{equation}
where $U\left(\mathbf{R}\right)=E$ is the potential energy. Because the functional $\Omega\left[V\right]$ is convex, its stationary point corresponds to the global minimum, which is obtained when the variational derivative with respect to $V$ vanishes and is given by
\begin{equation}
V(\mathbf{s}) = -F(\mathbf{s}) - \frac{1}{\beta} \ln p(\mathbf{s}).
\end{equation}

Thus, modifying the Hamiltonian becomes an optimization problem defined by the target distribution $p\left(\mathbf{s}\right)$. By selecting appropriate CVs, one can perform a VES simulation to generate an MTMB ensemble over the desired pressure and temperature intervals. In addition to the potential energy $E$ and volume $V$, we employ the structure factor $s_x$ as a CV, since it is a well-established indicator of first-order phase transitions in SiO$_2$ \cite{niu2018}, and has also proven effective in more complex systems \cite{deng2023}. The structure factor $s_x$ is expressed as a descriptor, 
\begin{equation}
s_x = s_{222_{Si}}^{3D}.
\end{equation}
This descriptor follows the Debye form of the structure factor and can be calculated using the Debye scattering function, expressed as:
\begin{equation}
s_{hkl}^{3D} = \frac{1}{N} \sum_{i=1}^N \sum_{j=1}^N 
f_i(Q) f_j(Q) \,
\frac{\sin \big( Q \cdot R_{ij} \big)}{Q \cdot R_{ij}} \,
w(R_{ij}),
\end{equation}
where $hkl$ denotes the Miller indices, $Q$ is the scattering vector, $f\left(Q\right)$ is the atomic scattering factor, and $R_{ij}$ is the distance between atoms $i$ and $j$. To reduce artifacts from the finite simulation box, a window function $w(R_{ij})$ is applied to smooth the sharp cutoff $R_c = 16 \,\text{\AA}$, defined as
\begin{equation}
w(R_{ij}) = \frac{\sin\!\left( Q \cdot R_{ij} / R_c \right)}{Q \cdot R_{ij} / R_c}.
\end{equation}

The descriptor $s_{222_{Si}}^{3D}$ corresponds to the first main peak of the structure factor intensitiy of Si atoms. This enables quantitative assessment of the similarity between an arbitrary structure and each phase, thereby effectively distinguishing among seifertite, pyrite-type, and liquid phases. We performed MTMB MD simulations on SiO$_2$ systems with 96 atoms using LAMMPS \cite{plimpton1995, thompson2022, zeng2025} augmented with PLUMED 2 \cite{tribello2014}. To cover pressures of 100–400 GPa and temperatures of 1000–10000 K, we expanded the bias potential in Legendre polynomials defined over the ranges $-92000<E<-51000$  kJ/mol, $440<V<740$ Å$^3$, and $200<s_x<450$. The energy threshold $\epsilon/\beta$ was set to $100 k_BT$ to facilitate sampling of first-order phase transitions. Using MTMB enhanced sampling and iterative training method \cite{deng2023}, we efficiently constructed MLPs for SiO$_2$ that cover a wide pressure and temperature range of 100–400 GPa and 1000–10000 K with only 7598 configurations.

\subsection{DeePMD framework}
The DeePMD approach employs deep neural networks to learn atomic environment representations and their direct mapping to potential energy \cite{zhang2018, zeng2025}. In this approach, the local environment of each atom is encoded from the relative positions of neighboring atoms within a cutoff radius, expressed in a local coordinate frame that preserves rotational, translational, and permutational invariance. A descriptor network converts these local configurations into symmetry-invariant vectors, which are then passed to a separate fitting network that maps them onto atomic energy contributions. The total potential energy is obtained by summing all atomic energies, while forces and virials are computed analytically as derivatives of the energy with respect to atomic positions and cell parameters. In this study, we used a descriptor network with three layers containing 25, 50, and 100 nodes, respectively, and a fitting network with three layers of 240 nodes each. A 6 Å cutoff distance ensured complete capture of the local atomic environments.

\subsection{\textit{Ab initio} calculations}
We performed \textit{ab initio} calculations based on the SCAN \cite{sun2015} and PBEsol \cite{perdew2008} XC functionals using VASP \cite{kresse1996}, to obtain energies, forces, and stresses of selected configurations sampled from the MTMB MD simulations. This procedure distinguishes our two MLPs, namely the one trained on data from the SCAN functional (MLP-SCAN) and the other from the PBEsol functional (MLP-PBEsol). The projector augmented wave (PAW) method \cite{kresse1999} was used as implemented in VASP \cite{kresse1996}. The core radii are $0.820$ Å for O and $1.312$ Å for Si, corresponding to $2s^2{2p}^4$ and $3s^2{3p}^2$ valence electron configurations, respectively. To achieve high precision, the energy cutoff of 800 eV was used to set the size of the basis set. The convergence error for the self-consistent solution to the Kohn-Sham equations was set to $10^{-6}$ eV, and the Brillouin zone was sampled using a $2\times2\times2$ Monkhorst-Pack mesh. This level of accuracy proved crucial for obtaining the training dataset needed to build a reliable MLP covering a broad region of the SiO$_2$ phase diagram (Fig. S1).

\subsection{Two-phase simulations}
To determine the melting temperatures of seifertite and pyrite-type SiO$_2$, we conducted two-phase coexistence simulations using LAMMPS \cite{plimpton1995, thompson2022} interfaced with DeePMD-kit \cite{zeng2025}. In this approach, solid and liquid regions are initially placed in direct contact within the simulation cell. During equilibration, the system naturally evolves toward either complete melting or complete crystallization, enabling the melting temperature to be identified from the condition of long-term phase stability. 

We first constructed $6\times12\times6$ supercells (5184 atoms) for both seifertite and pyrite-type and relaxed them for 10 ps with a 1 fs timestep under the isothermal-isobaric ensemble (NPT) at the target pressure-temperature conditions. To create an initial 1:1 solid-liquid configuration, atoms in half of each relaxed supercell were fixed while the other half was heated to a temperature well above the expected melting point in the canonical ensemble (NVT) for 5 ps. The resulting half-molten and half-crystalline structures were then equilibrated for 1 ps at the target conditions to serve as starting configurations. 

The two-phase coexistence simulations were conducted in the NPT ensemble at the desired pressure and temperature conditions. A fully molten (or crystalline) final structure after the simulations indicates that the simulation temperature is above (or below) the melting point. Each run lasted at least 500 ps and up to 1 ns to ensure convergence to a single phase. For both seifertite and pyrite-type, simulations were performed at pressure intervals of 20 GPa. The temperature interval was adaptively refined, starting from 200 K and decreasing to 10 K, to minimize computational cost while reducing uncertainty in the derived melting temperatures.

\section{Results and Discussion}
\subsection{Validation of machine learning potential}

\begin{figure*}[t]
\centering
    \includegraphics[width=0.90\textwidth]{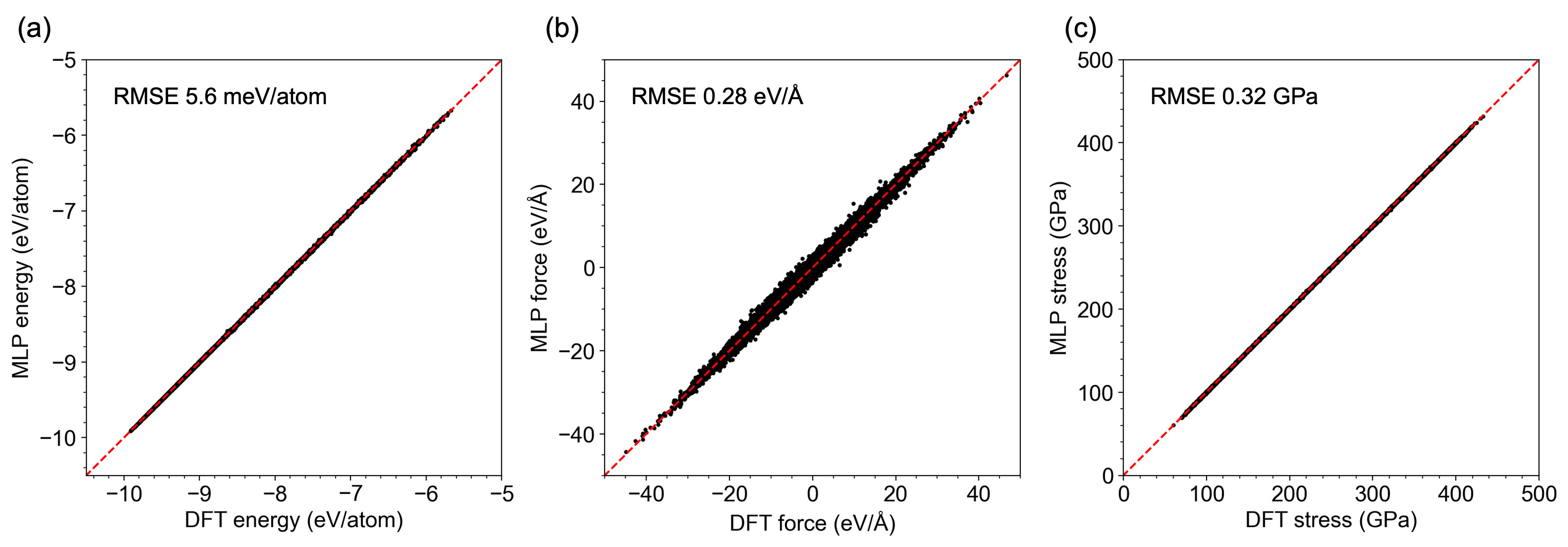}
\caption{\label{fig:1}Comparison between MLP-SCAN predictions and DFT calculations for energies (a), atomic forces (b), and stresses (c) using a test dataset of 10400 96-atom SiO$_2$ configurations over the temperature range 1000–10000 K and pressure range 100–400 GPa. The red dashed lines are given as guides for perfect matches.}
\end{figure*}

\begin{figure}[b]
\centering
    \includegraphics[width=0.48\textwidth]{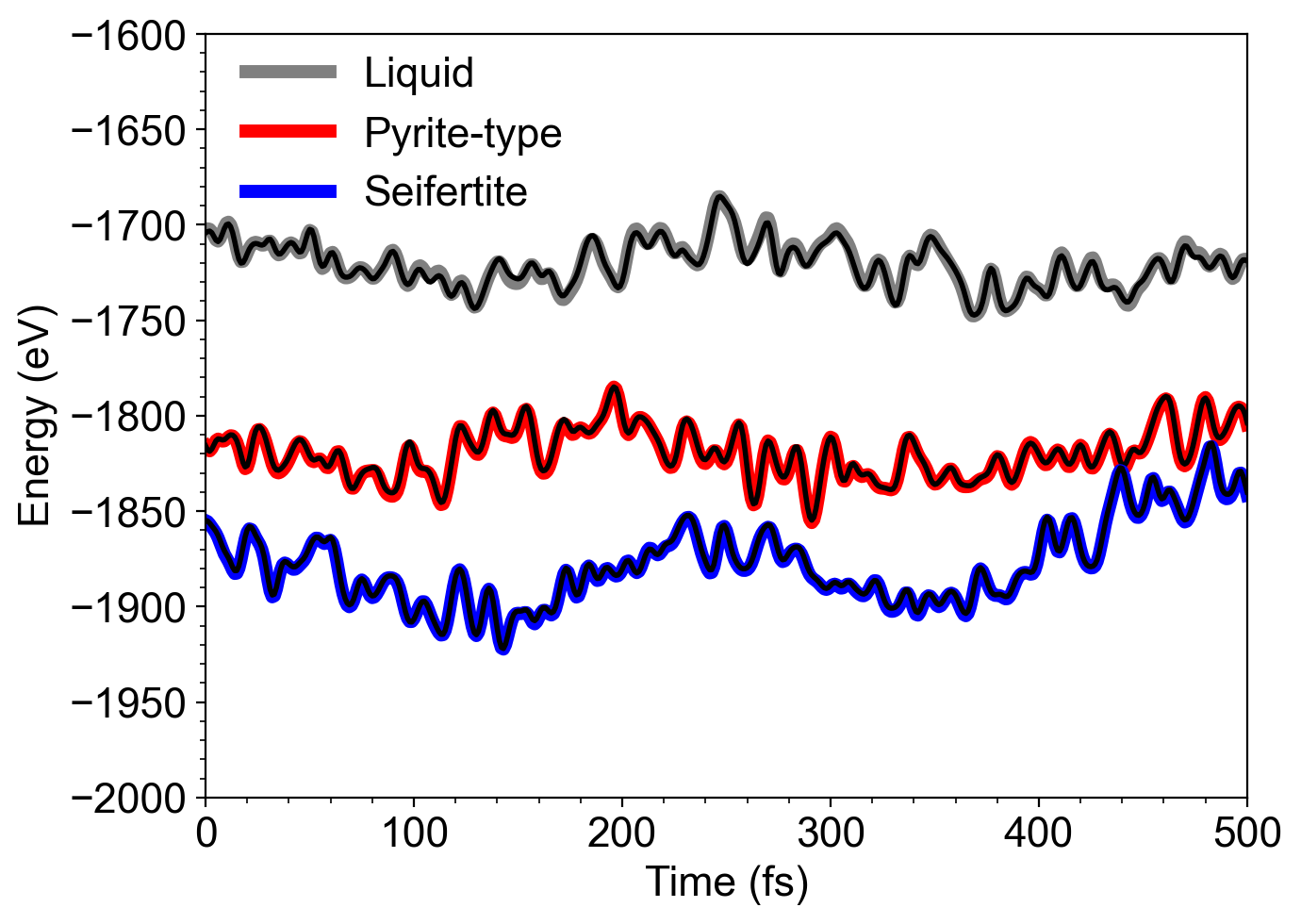}
\caption{\label{fig:2}Comparison between MLP-SCAN predictions (thin black lines) and DFT calculations (thick colored lines) of total energies for SiO$_2$ systems with 216 atoms at 200 GPa and 7000 K. None of the configurations in the trajectory were included in the training set. The root-mean-square errors of the MLP are 5.0, 5.0, and 6.5 meV/atom for seifertite, pyrite-type, and liquid, respectively.}
\end{figure}

To evaluate the accuracy of the two MLPs, we compared their predictions of energies, forces, and stresses with DFT results for 10400 configurations not included in the training dataset. The root-mean-square errors (RMSEs) of the energy, atomic force, and stress were 5.6 meV/atom, 0.28 eV/Å, and 0.32 GPa for MLP-SCAN (Fig.~\ref{fig:1}), and 9.4 meV/atom, 0.35 eV/Å, and 0.60 GPa for MLP-PBEsol (Fig. S2), respectively. These uncertainties are comparable to the precision of conventional AIMD simulations \cite{deng2021}. We further validated transferability by testing on larger systems containing 216 atoms, which were not included in the training set, as all training configurations contained 96 atoms. The RMSEs of energy predictions for these larger systems ranged from 5.0 to 6.5 meV/atom with MLP-SCAN (Fig.~\ref{fig:2}) and from 7.9 to 9.8 meV/atom with MLP-PBEsol (Fig. S3), consistent with the 96-atom results and thus validating both the accuracy and transferability of these MLPs to larger systems.

\subsection{Melting properties of seifertite and pyrite-type}
\begin{figure*}[t]
\centering
    \includegraphics[width=0.90\textwidth]{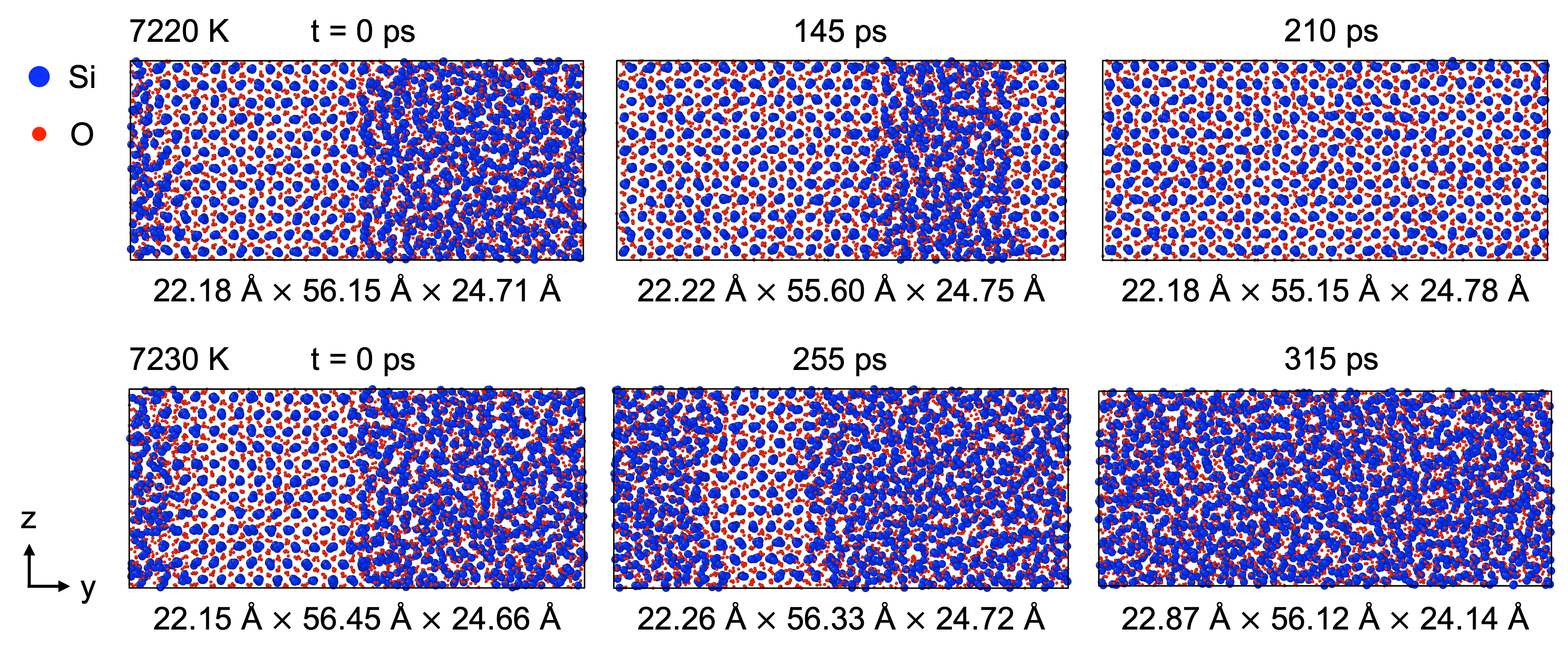}
\caption{\label{fig:3}Two-phase simulations of seifertite SiO$_2$ and liquid coexistence using the machine learning potential based on the SCAN functional at 200 GPa and 7220 K (upper panel) and 7230 K (lower panel). The simulation box contains 1728 SiO$_2$ formula units (5184 atoms). The box centers are evenly spaced horizontally and vertically aligned to clearly illustrate the relative volume differences.}
\end{figure*}

\begin{figure*}
\centering
    \includegraphics[width=0.90\textwidth]{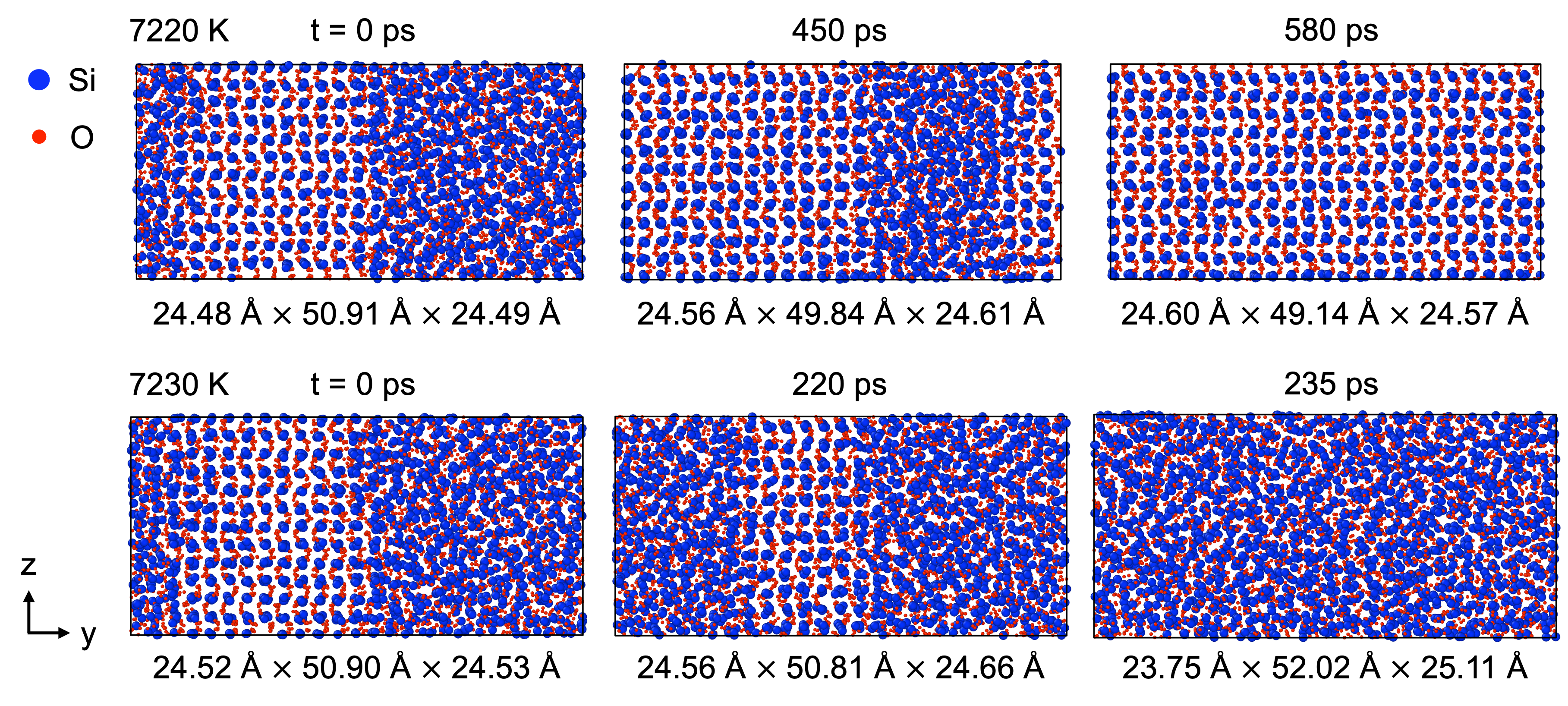}
\caption{\label{fig:4}Two-phase simulations of pyrite-type SiO$_2$ and liquid coexistence using the machine learning potential based on the SCAN functional 200 GPa and 7220 K (upper panel) and 7230 K (lower panel). The simulation box contains 1728 SiO$_2$ formula units (5184 atoms). The box centers are evenly spaced horizontally and vertically aligned to clearly illustrate the relative volume differences.}
\end{figure*}

\begin{figure*}[t]
\centering
    \includegraphics[width=1.0\textwidth]{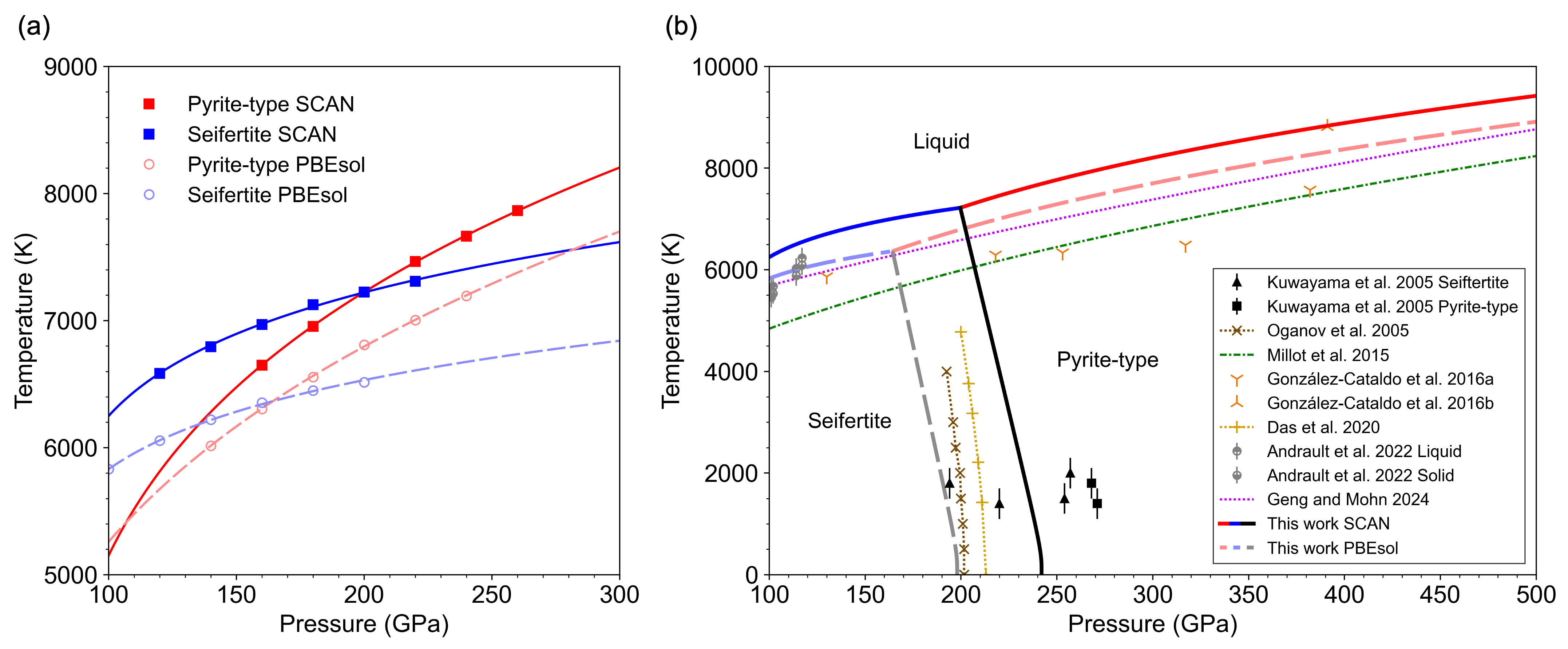}
\caption{\label{fig:5}Melting phase relation of seifertite and pyrite-type SiO$_2$. Melting points determined by two-phase simulations were fitted and extrapolated using the Simon equation (a). The resulting melting curves, along with the solid-solid phase boundary between seifertite and pyrite-type, were plotted together with results from the literature (b). Experimental data points are shown with filled markers \cite{kuwayama2005} or half-filled markers \cite{andrault2022}, with error bars indicating temperature uncertainty. One shock compression experiment is represented separately by a green dashed-dotted line \cite{millot2015}.}
\end{figure*}

\begin{table*}[t]
\caption{Melting thermodynamic properties of SiO$_2$: pressure $P$, melting temperature $T_m$, slope of the melting curve $dT/dP$, volume of melting $\Delta V_m$, entropy of melting $\Delta S_m$, and enthalpy of melting $\Delta H_m$.}
\begin{ruledtabular}
\begin{tabular}{l c *{5}{c c}}
Phase 
& $P$ (GPa)
& \multicolumn{2}{c}{$T_m$ (K)} 
& \multicolumn{2}{c}{$dT/dP$ (K/GPa)} 
& \multicolumn{2}{c}{$\Delta V_m$ (\AA$^{3}$/atom)} 
& \multicolumn{2}{c}{$\Delta S_m$ (J mol$^{-1}$ K$^{-1}$)} 
& \multicolumn{2}{c}{$\Delta H_m$ (kJ mol$^{-1}$)} \\
\cline{3-4}\cline{5-6}\cline{7-8}\cline{9-10}\cline{11-12}
&
& SCAN & PBEsol 
& SCAN & PBEsol 
& SCAN & PBEsol 
& SCAN & PBEsol 
& SCAN & PBEsol \\
\hline
Seifertite 
& 100 &   -- & 5830 &    -- & 14.14 &    -- & 0.303 &   -- & 38.7 &  -- & 226\\
& 120 & 6585 & 6055 & 13.13 &  9.29 & 0.294 & 0.227 & 40.4 & 44.2 & 266 & 268\\
& 140 & 6795 & 6220 &  9.45 &  6.97 & 0.229 & 0.172 & 43.8 & 44.5 & 298 & 277\\
& 160 & 6970 & 6355 &  7.42 &  5.60 & 0.182 & 0.126 & 44.2 & 40.6 & 308 & 258\\
& 180 & 7125 & 6450 &  6.13 &  4.70 & 0.142 & 0.088 & 41.8 & 34.0 & 298 & 219\\
& 200 & 7225 & 6515 &  5.24 &  4.05 & 0.108 & 0.058 & 37.4 & 25.9 & 270 & 169\\
& 220 & 7310 &   -- &  4.58 &    -- & 0.080 &    -- & 31.6 &   -- & 231 &  --\\
\hline
Pyrite-type
& 140 &   -- & 6015 &    -- & 15.70 &    -- & 0.319 &   -- & 36.7 &  -- & 221\\
& 160 & 6650 & 6305 & 16.73 & 13.66 & 0.319 & 0.276 & 34.4 & 36.5 & 229 & 230\\
& 180 & 6955 & 6555 & 14.24 & 12.15 & 0.283 & 0.243 & 36.0 & 36.1 & 250 & 237\\
& 200 & 7225 & 6810 & 12.46 & 10.98 & 0.254 & 0.215 & 36.9 & 35.4 & 266 & 241\\
& 220 & 7465 & 7005 & 11.11 & 10.04 & 0.228 & 0.191 & 37.1 & 34.3 & 277 & 240\\
& 240 & 7665 & 7195 & 10.06 &  9.27 & 0.204 & 0.172 & 36.7 & 33.6 & 281 & 242\\
& 260 & 7865 &   -- &  9.21 &    -- & 0.184 &    -- & 36.2 &   -- & 285 &  --\\
\end{tabular}
\end{ruledtabular}
\label{tab:1}
\end{table*}

From the two-phase coexistence simulations, we precisely determined the melting curves of both seifertite and pyrite-type. At 200 GPa, seifertite completely crystallized after 210 ps at 7220 K, whereas it fully melted after 315 ps at 7230 K (Fig.~\ref{fig:3}), yielding a melting temperature of 7225 K with 5 K uncertainty. Pyrite-type exhibited identical crystallization and melting behavior as seifertite at 200 GPa (Fig.~\ref{fig:4}), establishing the seifertite/pyrite-type/liquid triple point near 7225 K. The melting points of both phases were fitted to the Simon equations [Table~\ref{tab:1}, Fig.~\ref{fig:5}(a)]. The melting curve of seifertite can be expressed as 
$T_{m}=(6585\pm5)(\frac{P-120}{45.77\pm7.09}+1)^{\frac{1}{10.96\pm1.10}}$ using MLP-SCAN, and
$T_{m}=(5830\pm5)(\frac{P-100}{34.36\pm5.02}+1)^{\frac{1}{12.00\pm1.05}}$ using MLP-PBEsol, where $T_{m}$ is the melting temperature (K) and $P$ is the pressure (GPa).
Similarly, for pyrite-type, the melting curve can also be expressed as
$T_{m}=(6650\pm5)(\frac{P-160}{87.15\pm3.91}+1)^{\frac{1}{4.56\pm0.15}}$ using MLP-SCAN, and
$T_{m}=(6015\pm5)(\frac{P-140}{97.42\pm9.52}+1)^{\frac{1}{3.93\pm0.29}}$ using MLP-PBEsol. 

Our melting curve obtained using MLP-PBEsol agrees well with the recent PBE-based theoretical study \cite{geng2024} (Fig.~\ref{fig:5}). Its slope is broadly consistent with shock-compression measurements \cite{millot2015}, although our absolute melting temperatures are about 1000 K higher. Both MLP-SCAN and MLP-PBEsol show similar slopes without abrupt changes, except near the triple point where the slope nearly doubles. MLP-SCAN predicts melting temperatures 6–10 \% higher than MLP-PBEsol. It has been shown that SCAN achieves significantly improved accuracy in predicting lattice parameters \cite{sun2015}, formation energies \cite{sun2016}, lattice dynamics \cite{ning2022}, and melting temperatures \cite{rang2019,jinnouchi2019} across a wide range of materials with diverse bonding characteristics, owing to its ability to capture intermediate-range van der Waals interactions that are largely neglected by GGA functionals such as PBEsol. Notably, for MgO, which is an oxide with partial covalent bonding similar to that of SiO$_2$, SCAN predicts a melting temperature of 3032 K at 1 bar that is in excellent agreement with the experimental range (3040–3250 K) and substantially improves upon the PBE result (2747 K) \cite{rang2019}. Therefore, we expect that SCAN also captures the structural and thermodynamic properties of SiO$_2$ more accurately than PBEsol, with more accurate melting temperature estimates. At 120 GPa, MLP-SCAN yields a melting temperature of 6585 K, about 6 \% higher than laser-heated DAC measurements \cite{andrault2020,andrault2022}, with the actual discrepancy likely even smaller when considering the experimental uncertainties \cite{deng2017}.

\begin{figure*}[t]
\centering
    \includegraphics[width=0.75\textwidth]{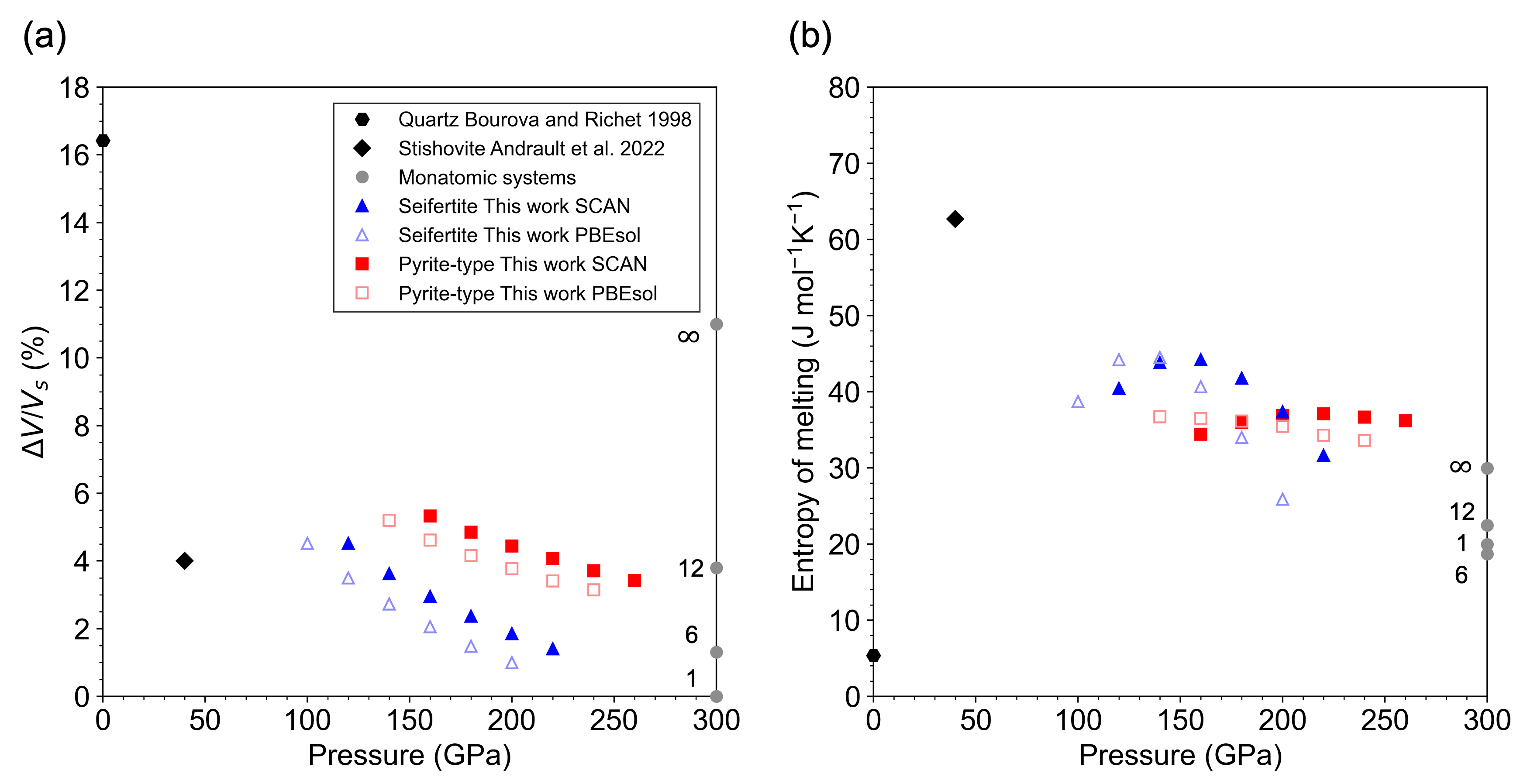}
\caption{\label{fig:6}The volume of melting (a) and the entropy of melting (b) from the seifertite and pyrite-type phases. For comparison, experimental values (black markers) for quartz \cite{bourova1998} and stishovite \cite{andrault2022}, along with theoretical results for monatomic systems interacting via repulsive inverse-power potentials (gray circles) \cite{young1991}, are also shown. The power-law exponent ranges from 1 (one-component plasma) to infinity (hard spheres).
}
\end{figure*}

We also tracked the contraction of the simulation box upon crystallization and its expansion upon melting. From these changes, we obtained the volume of melting ($\Delta V_m$) and, using the Clapeyron slope ($\gamma = dP/dT_m$) derived from the melting curve, the entropy of melting ($\Delta S_m = \gamma\Delta V_m$) (Table~\ref{tab:1}, Fig.~\ref{fig:6}). The enthalpy of melting was subsequently determined as $\Delta H_m = T_m \Delta S_m$ (Table~\ref{tab:1}). For both seifertite and pyrite-type phases, $\Delta V_m$ decreases with increasing pressure owing to the greater compressibility of the liquid relative to the solid, and remains positive across all conditions.

The entropies of melting for both phases are larger than $Rln2$ (where $R$ is the gas constant) and those of simple monatomic liquids with inverse-power repulsive interactions. This larger $\Delta S_m$ reflects the diverse Si-O coordination environments in the melt, which give rise to structural complexity absent from the nearly close-packed monatomic liquids. For both phases,  $\Delta S_m$ ranges from 30 to 45 J mol$^{-1}$ K$^{-1}$, and only vary moderately with compression. The melting entropy values are between those of quartz at ambient pressure \cite{bourova1998} and stishovite at 40 GPa \cite{andrault2022}  (Fig.~\ref{fig:6}).

\subsection{Seifertite to pyrite-type transition}
To determine the solid-solid phase boundary between seifertite and pyrite-type, we first identified the triple point and the phase transition pressure at 0 K. The intersection of the seifertite and pyrite-type melting curves yields the seifertite/pyrite-type/liquid triple point at 200 GPa and 7220 K for the MLP-SCAN model, and at 165 GPa and 6370 K for MLP-PBEsol (Fig.~\ref{fig:5}). The 0 K transition pressure was determined from DFT calculations by locating the enthalpy crossover ($\Delta H=H_{pyrite}-H_{seifertite}=0$). SCAN predicts a phase transition at 242 GPa, approximately 22 \% higher than the 198 GPa obtained with PBEsol (Fig. S4). Combining these results with the calculated volume difference between the two phases (0.493 and 0.542 Å$^3$/formula unit for SCAN and PBEsol, respectively) and the Debye temperature of seifertite (1133 K) \cite{stixrude2024}, and applying the thermodynamic formalism of Jeanloz \cite{knittle1989}, we derived the solid-solid phase boundary between seifertite and pyrite-type (Fig.~\ref{fig:5}).

The phase boundary obtained with the SCAN functional shows much closer agreement with the experimental estimate of $\sim$260 GPa \cite{kuwayama2005} than with previous theoretical predictions \cite{oganov2005,das2020,tsuchiya2004}. Notably, Kuwayama \textit{et al.} \cite{kuwayama2005} employed the equation of state (EOS) of platinum (Pt) by Holmes \textit{et al.} \cite{holmes1989}, which overestimates pressure by up to $\sim$8 \% in that pressure range compared with more recent Pt EOS determinations \cite{fei2007,yokoo2009,matsui2009,dorfman2012} (Fig. S5). Correcting for this systematic offset places the experimental phase boundary between seifertite and pyrite-type at lower pressures, around 240–255 GPa at temperatures below 2000 K. Therefore, our solid-solid phase boundary based on the SCAN functional, which yields a 0 K transition pressure of 242 GPa, shows even better agreement with previous experimental results when the updated Pt EOS calibrations are considered.

Both phase boundaries obtained with MLP-SCAN and MLP-PBEsol exhibit negative Clapeyron slopes, approximately –6.1 MPa/K for SCAN and –5.5 MPa/K for PBEsol (Fig.~\ref{fig:5}). The negative Clapeyron slope ($dP/dT=\Delta S/\Delta V$) reflects a positive entropy change ($\Delta S>0$) combined with a volume decrease ($\Delta V<0$) of $\sim$2.5 \% at the triple point and $\sim$3.1 \% at 0 K as seifertite transforms to pyrite-type, for both functionals. Previous theoretical studies reported Clapeyron slopes near –2.8 MPa/K for the seifertite to pyrite-type phase boundary \cite{oganov2005,das2020}, roughly half of our value. This discrepancy likely stems from the quasi-harmonic approximation (QHA) used in those works, which struggles to capture anharmonic effects and defect contributions that affect free energy and phase boundary predictions at high temperatures \cite{wu2009}. By contrast, our two-phase MD simulations inherently include anharmonicity and defect dynamics through direct sampling of a large system of 5184 atoms. The approximately twofold steeper negative Clapeyron slope in our results suggests stronger stratification within the mantles of super-Earth exoplanets, thereby promoting layered convection.
Latent heat absorption locally reduces thermal buoyancy, and a strongly negative slope amplifies this effect, making the boundary more resistant to vertical flow and acting as a dynamical barrier \cite{schubert1975,faccenda2017}. As a result, such a barrier may slow the secular cooling of the mantle \cite{christensen1995} and weaken the generation of magnetic fields in super-Earths \cite{christensen2018}, thereby affecting their long-term habitability potential \cite{vanhoolst2019}.

\section{Conclusion}
We developed two machine learning potentials for SiO$_2$ that accurately capture interatomic interactions across multiple phases and a wide range of pressure (100–400 GPa) and temperature (1000–10000 K) conditions, using both the SCAN meta-GGA and PBEsol GGA exchange-correlation functionals. Combining these two MLPs with two-phase coexistence simulations, we determined the melting curve and the seifertite to pyrite-type phase boundary, as well as key thermodynamic properties of melting, including the Clapeyron slope, entropy of melting, volume of melting, and enthalpy of melting.

The SCAN-based MLP yields the highest melting temperatures reported, 6–10\% higher than those predicted by the PBEsol-based MLP. Given that SCAN accounts for intermediate-range interactions that LDA and GGA largely miss, it likely offers a more accurate description of melting behavior. The SCAN functional also predicts a seifertite to pyrite-type phase transition boundary that most closely matches experimental observations.. The strongly negative Clapeyron slope of –6.1 MPa/K at the seifertite to pyrite-type transition suggests that mantle convection could be layered in super-Earths, which may in turn lead to sluggish planetary cooling and a weak magnetic field.

\begin{acknowledgments}
J.D. acknowledges support from the National Science Foundation (Grant EAR-2444522). X.D. is supported by the Carnegie Postdoctoral Fellowship. All computations for this work were carried out on the Della and Tiger clusters managed and supported by Princeton University's Research Computing.
\end{acknowledgments}

\section*{Data Availability}
The training and test datasets used to construct and validate the machine learning potentials in this study have been deposited at the Open Science Framework and are available online \cite{osf_y9qnm}.

\bibliography{manuscript}

\begin{thebibliography}{5}%
\makeatletter
\providecommand \@ifxundefined [1]{%
 \@ifx{#1\undefined}
}%
\providecommand \@ifnum [1]{%
 \ifnum #1\expandafter \@firstoftwo
 \else \expandafter \@secondoftwo
 \fi
}%
\providecommand \@ifx [1]{%
 \ifx #1\expandafter \@firstoftwo
 \else \expandafter \@secondoftwo
 \fi
}%
\providecommand \natexlab [1]{#1}%
\providecommand \enquote  [1]{``#1''}%
\providecommand \bibnamefont  [1]{#1}%
\providecommand \bibfnamefont [1]{#1}%
\providecommand \citenamefont [1]{#1}%
\providecommand \href@noop [0]{\@secondoftwo}%
\providecommand \href [0]{\begingroup \@sanitize@url \@href}%
\providecommand \@href[1]{\@@startlink{#1}\@@href}%
\providecommand \@@href[1]{\endgroup#1\@@endlink}%
\providecommand \@sanitize@url [0]{\catcode `\\12\catcode `\$12\catcode `\&12\catcode `\#12\catcode `\^12\catcode `\_12\catcode `\%12\relax}%
\providecommand \@@startlink[1]{}%
\providecommand \@@endlink[0]{}%
\providecommand \url  [0]{\begingroup\@sanitize@url \@url }%
\providecommand \@url [1]{\endgroup\@href {#1}{\urlprefix }}%
\providecommand \urlprefix  [0]{URL }%
\providecommand \Eprint [0]{\href }%
\providecommand \doibase [0]{https://doi.org/}%
\providecommand \selectlanguage [0]{\@gobble}%
\providecommand \bibinfo  [0]{\@secondoftwo}%
\providecommand \bibfield  [0]{\@secondoftwo}%
\providecommand \translation [1]{[#1]}%
\providecommand \BibitemOpen [0]{}%
\providecommand \bibitemStop [0]{}%
\providecommand \bibitemNoStop [0]{.\EOS\space}%
\providecommand \EOS [0]{\spacefactor3000\relax}%
\providecommand \BibitemShut  [1]{\csname bibitem#1\endcsname}%
\let\auto@bib@innerbib\@empty
\bibitem [{\citenamefont {Holmes}\ \emph {et~al.}(1989)\citenamefont {Holmes}, \citenamefont {Moriarty}, \citenamefont {Gathers},\ and\ \citenamefont {Nellis}}]{holmes1989}%
  \BibitemOpen
  \bibfield  {author} {\bibinfo {author} {\bibfnamefont {N.~C.}\ \bibnamefont {Holmes}}, \bibinfo {author} {\bibfnamefont {J.~A.}\ \bibnamefont {Moriarty}}, \bibinfo {author} {\bibfnamefont {G.~R.}\ \bibnamefont {Gathers}},\ and\ \bibinfo {author} {\bibfnamefont {W.~J.}\ \bibnamefont {Nellis}},\ }\bibfield  {title} {\bibinfo {title} {The equation of state of platinum to 660 gpa (6.6 mbar)},\ }\href {https://doi.org/10.1063/1.344047} {\bibfield  {journal} {\bibinfo  {journal} {J.\ Appl.\ Phys.}\ }\textbf {\bibinfo {volume} {66}},\ \bibinfo {pages} {2962} (\bibinfo {year} {1989})}\BibitemShut {NoStop}%
\bibitem [{\citenamefont {Fei}\ \emph {et~al.}(2007)\citenamefont {Fei}, \citenamefont {Ricolleau}, \citenamefont {Frank}, \citenamefont {Mibe}, \citenamefont {Shen},\ and\ \citenamefont {Prakapenka}}]{fei2007}%
  \BibitemOpen
  \bibfield  {author} {\bibinfo {author} {\bibfnamefont {Y.}~\bibnamefont {Fei}}, \bibinfo {author} {\bibfnamefont {A.}~\bibnamefont {Ricolleau}}, \bibinfo {author} {\bibfnamefont {M.}~\bibnamefont {Frank}}, \bibinfo {author} {\bibfnamefont {K.}~\bibnamefont {Mibe}}, \bibinfo {author} {\bibfnamefont {G.}~\bibnamefont {Shen}},\ and\ \bibinfo {author} {\bibfnamefont {V.}~\bibnamefont {Prakapenka}},\ }\bibfield  {title} {\bibinfo {title} {Toward an internally consistent pressure scale},\ }\href {https://doi.org/10.1073/pnas.0609013104} {\bibfield  {journal} {\bibinfo  {journal} {Proc.\ Natl.\ Acad.\ Sci.}\ }\textbf {\bibinfo {volume} {104}},\ \bibinfo {pages} {9182} (\bibinfo {year} {2007})}\BibitemShut {NoStop}%
\bibitem [{\citenamefont {Yokoo}\ \emph {et~al.}(2009)\citenamefont {Yokoo}, \citenamefont {Kawai}, \citenamefont {Nakamura}, \citenamefont {i.~Kondo}, \citenamefont {Tange},\ and\ \citenamefont {Tsuchiya}}]{yokoo2009}%
  \BibitemOpen
  \bibfield  {author} {\bibinfo {author} {\bibfnamefont {M.}~\bibnamefont {Yokoo}}, \bibinfo {author} {\bibfnamefont {N.}~\bibnamefont {Kawai}}, \bibinfo {author} {\bibfnamefont {K.~G.}\ \bibnamefont {Nakamura}}, \bibinfo {author} {\bibfnamefont {K.}~\bibnamefont {i.~Kondo}}, \bibinfo {author} {\bibfnamefont {Y.}~\bibnamefont {Tange}},\ and\ \bibinfo {author} {\bibfnamefont {T.}~\bibnamefont {Tsuchiya}},\ }\bibfield  {title} {\bibinfo {title} {Ultrahigh-pressure scales for gold and platinum at pressures up to 550 gpa},\ }\href {https://doi.org/10.1103/PhysRevB.80.104114} {\bibfield  {journal} {\bibinfo  {journal} {Phys.\ Rev.\ B}\ }\textbf {\bibinfo {volume} {80}},\ \bibinfo {pages} {104114} (\bibinfo {year} {2009})}\BibitemShut {NoStop}%
\bibitem [{\citenamefont {Dorfman}\ \emph {et~al.}(2012)\citenamefont {Dorfman}, \citenamefont {Prakapenka}, \citenamefont {Meng},\ and\ \citenamefont {Duffy}}]{dorfman2012}%
  \BibitemOpen
  \bibfield  {author} {\bibinfo {author} {\bibfnamefont {S.~M.}\ \bibnamefont {Dorfman}}, \bibinfo {author} {\bibfnamefont {V.~B.}\ \bibnamefont {Prakapenka}}, \bibinfo {author} {\bibfnamefont {Y.}~\bibnamefont {Meng}},\ and\ \bibinfo {author} {\bibfnamefont {T.~S.}\ \bibnamefont {Duffy}},\ }\bibfield  {title} {\bibinfo {title} {Intercomparison of pressure standards (au, pt, mo, mgo, nacl and ne) to 2.5 mbar},\ }\href {https://doi.org/10.1029/2012JB009292} {\bibfield  {journal} {\bibinfo  {journal} {J.\ Geophys.\ Res.}\ }\textbf {\bibinfo {volume} {117}},\ \bibinfo {pages} {B08210} (\bibinfo {year} {2012})}\BibitemShut {NoStop}%
\bibitem [{\citenamefont {Matsui}\ \emph {et~al.}(2009)\citenamefont {Matsui}, \citenamefont {Ito}, \citenamefont {Katsura}, \citenamefont {Yamazaki}, \citenamefont {Yoshino}, \citenamefont {Yokoyama},\ and\ \citenamefont {Funakoshi}}]{matsui2009}%
  \BibitemOpen
  \bibfield  {author} {\bibinfo {author} {\bibfnamefont {M.}~\bibnamefont {Matsui}}, \bibinfo {author} {\bibfnamefont {E.}~\bibnamefont {Ito}}, \bibinfo {author} {\bibfnamefont {T.}~\bibnamefont {Katsura}}, \bibinfo {author} {\bibfnamefont {D.}~\bibnamefont {Yamazaki}}, \bibinfo {author} {\bibfnamefont {T.}~\bibnamefont {Yoshino}}, \bibinfo {author} {\bibfnamefont {A.}~\bibnamefont {Yokoyama}},\ and\ \bibinfo {author} {\bibfnamefont {K.}~\bibnamefont {Funakoshi}},\ }\bibfield  {title} {\bibinfo {title} {The temperature-pressure-volume equation of state of platinum},\ }\href {https://doi.org/10.1063/1.3054331} {\bibfield  {journal} {\bibinfo  {journal} {J.\ Appl.\ Phys.}\ }\textbf {\bibinfo {volume} {105}},\ \bibinfo {pages} {013505} (\bibinfo {year} {2009})}\BibitemShut {NoStop}%
\end{thebibliography}%


\begin{thebibliography}{54}%
\makeatletter
\providecommand \@ifxundefined [1]{%
 \@ifx{#1\undefined}
}%
\providecommand \@ifnum [1]{%
 \ifnum #1\expandafter \@firstoftwo
 \else \expandafter \@secondoftwo
 \fi
}%
\providecommand \@ifx [1]{%
 \ifx #1\expandafter \@firstoftwo
 \else \expandafter \@secondoftwo
 \fi
}%
\providecommand \natexlab [1]{#1}%
\providecommand \enquote  [1]{``#1''}%
\providecommand \bibnamefont  [1]{#1}%
\providecommand \bibfnamefont [1]{#1}%
\providecommand \citenamefont [1]{#1}%
\providecommand \href@noop [0]{\@secondoftwo}%
\providecommand \href [0]{\begingroup \@sanitize@url \@href}%
\providecommand \@href[1]{\@@startlink{#1}\@@href}%
\providecommand \@@href[1]{\endgroup#1\@@endlink}%
\providecommand \@sanitize@url [0]{\catcode `\\12\catcode `\$12\catcode `\&12\catcode `\#12\catcode `\^12\catcode `\_12\catcode `\%12\relax}%
\providecommand \@@startlink[1]{}%
\providecommand \@@endlink[0]{}%
\providecommand \url  [0]{\begingroup\@sanitize@url \@url }%
\providecommand \@url [1]{\endgroup\@href {#1}{\urlprefix }}%
\providecommand \urlprefix  [0]{URL }%
\providecommand \Eprint [0]{\href }%
\providecommand \doibase [0]{https://doi.org/}%
\providecommand \selectlanguage [0]{\@gobble}%
\providecommand \bibinfo  [0]{\@secondoftwo}%
\providecommand \bibfield  [0]{\@secondoftwo}%
\providecommand \translation [1]{[#1]}%
\providecommand \BibitemOpen [0]{}%
\providecommand \bibitemStop [0]{}%
\providecommand \bibitemNoStop [0]{.\EOS\space}%
\providecommand \EOS [0]{\spacefactor3000\relax}%
\providecommand \BibitemShut  [1]{\csname bibitem#1\endcsname}%
\let\auto@bib@innerbib\@empty
\bibitem [{\citenamefont {Kuwayama}\ \emph {et~al.}(2005)\citenamefont {Kuwayama}, \citenamefont {Hirose}, \citenamefont {Sata},\ and\ \citenamefont {Ohishi}}]{kuwayama2005}%
  \BibitemOpen
  \bibfield  {author} {\bibinfo {author} {\bibfnamefont {Y.}~\bibnamefont {Kuwayama}}, \bibinfo {author} {\bibfnamefont {K.}~\bibnamefont {Hirose}}, \bibinfo {author} {\bibfnamefont {N.}~\bibnamefont {Sata}},\ and\ \bibinfo {author} {\bibfnamefont {Y.}~\bibnamefont {Ohishi}},\ }\bibfield  {title} {\bibinfo {title} {The pyrite-type high-pressure form of silica},\ }\href {https://doi.org/10.1126/science.1114879} {\bibfield  {journal} {\bibinfo  {journal} {Science}\ }\textbf {\bibinfo {volume} {309}},\ \bibinfo {pages} {923} (\bibinfo {year} {2005})}\BibitemShut {NoStop}%
\bibitem [{\citenamefont {Oganov}\ \emph {et~al.}(2005)\citenamefont {Oganov}, \citenamefont {Gillan},\ and\ \citenamefont {Price}}]{oganov2005}%
  \BibitemOpen
  \bibfield  {author} {\bibinfo {author} {\bibfnamefont {A.~R.}\ \bibnamefont {Oganov}}, \bibinfo {author} {\bibfnamefont {M.~J.}\ \bibnamefont {Gillan}},\ and\ \bibinfo {author} {\bibfnamefont {G.~D.}\ \bibnamefont {Price}},\ }\bibfield  {title} {\bibinfo {title} {Structural stability of silica at high pressures and temperatures},\ }\href {https://doi.org/10.1103/PhysRevB.71.064104} {\bibfield  {journal} {\bibinfo  {journal} {Phys.\ Rev.\ B}\ }\textbf {\bibinfo {volume} {71}},\ \bibinfo {pages} {064104} (\bibinfo {year} {2005})}\BibitemShut {NoStop}%
\bibitem [{\citenamefont {Liu}\ \emph {et~al.}(2021)\citenamefont {Liu}, \citenamefont {Shi}, \citenamefont {Gao}, \citenamefont {Wang}, \citenamefont {Han}, \citenamefont {Lu}, \citenamefont {Wang}, \citenamefont {Xing},\ and\ \citenamefont {Sun}}]{liu2021}%
  \BibitemOpen
  \bibfield  {author} {\bibinfo {author} {\bibfnamefont {C.}~\bibnamefont {Liu}}, \bibinfo {author} {\bibfnamefont {J.}~\bibnamefont {Shi}}, \bibinfo {author} {\bibfnamefont {H.}~\bibnamefont {Gao}}, \bibinfo {author} {\bibfnamefont {J.}~\bibnamefont {Wang}}, \bibinfo {author} {\bibfnamefont {Y.}~\bibnamefont {Han}}, \bibinfo {author} {\bibfnamefont {X.}~\bibnamefont {Lu}}, \bibinfo {author} {\bibfnamefont {H.-T.}\ \bibnamefont {Wang}}, \bibinfo {author} {\bibfnamefont {D.}~\bibnamefont {Xing}},\ and\ \bibinfo {author} {\bibfnamefont {J.}~\bibnamefont {Sun}},\ }\bibfield  {title} {\bibinfo {title} {Mixed coordination silica at megabar pressure},\ }\href {https://doi.org/10.1103/PhysRevLett.126.035701} {\bibfield  {journal} {\bibinfo  {journal} {Phys.\ Rev.\ Lett.}\ }\textbf {\bibinfo {volume} {126}},\ \bibinfo {pages} {035701} (\bibinfo {year} {2021})}\BibitemShut {NoStop}%
\bibitem [{\citenamefont {Andrault}\ \emph {et~al.}(2020)\citenamefont {Andrault}, \citenamefont {Morard}, \citenamefont {Garbarino}, \citenamefont {Mezouar}, \citenamefont {Bouhifd},\ and\ \citenamefont {Kawamoto}}]{andrault2020}%
  \BibitemOpen
  \bibfield  {author} {\bibinfo {author} {\bibfnamefont {D.}~\bibnamefont {Andrault}}, \bibinfo {author} {\bibfnamefont {G.}~\bibnamefont {Morard}}, \bibinfo {author} {\bibfnamefont {G.}~\bibnamefont {Garbarino}}, \bibinfo {author} {\bibfnamefont {M.}~\bibnamefont {Mezouar}}, \bibinfo {author} {\bibfnamefont {M.~A.}\ \bibnamefont {Bouhifd}},\ and\ \bibinfo {author} {\bibfnamefont {T.}~\bibnamefont {Kawamoto}},\ }\bibfield  {title} {\bibinfo {title} {Melting behavior of sio$_2$ up to 120 gpa},\ }\href {https://doi.org/10.1007/s00269-019-01077-3} {\bibfield  {journal} {\bibinfo  {journal} {Phys.\ Chem.\ Miner.}\ }\textbf {\bibinfo {volume} {47}},\ \bibinfo {pages} {10} (\bibinfo {year} {2020})}\BibitemShut {NoStop}%
\bibitem [{\citenamefont {Andrault}\ \emph {et~al.}(2022)\citenamefont {Andrault}, \citenamefont {Pison}, \citenamefont {Morard}, \citenamefont {Garbarino}, \citenamefont {Mezouar}, \citenamefont {Bouhifd},\ and\ \citenamefont {Kawamoto}}]{andrault2022}%
  \BibitemOpen
  \bibfield  {author} {\bibinfo {author} {\bibfnamefont {D.}~\bibnamefont {Andrault}}, \bibinfo {author} {\bibfnamefont {L.}~\bibnamefont {Pison}}, \bibinfo {author} {\bibfnamefont {G.}~\bibnamefont {Morard}}, \bibinfo {author} {\bibfnamefont {G.}~\bibnamefont {Garbarino}}, \bibinfo {author} {\bibfnamefont {M.}~\bibnamefont {Mezouar}}, \bibinfo {author} {\bibfnamefont {M.~A.}\ \bibnamefont {Bouhifd}},\ and\ \bibinfo {author} {\bibfnamefont {T.}~\bibnamefont {Kawamoto}},\ }\bibfield  {title} {\bibinfo {title} {Comment on: Melting behavior of sio$_2$ up to 120 gpa (andrault et al. 2020)},\ }\href {https://doi.org/10.1007/s00269-021-01174-2} {\bibfield  {journal} {\bibinfo  {journal} {Phys.\ Chem.\ Miner.}\ }\textbf {\bibinfo {volume} {49}},\ \bibinfo {pages} {3} (\bibinfo {year} {2022})}\BibitemShut {NoStop}%
\bibitem [{\citenamefont {Millot}\ \emph {et~al.}(2015)\citenamefont {Millot}, \citenamefont {Dubrovinskaia}, \citenamefont {Černok}, \citenamefont {Blaha}, \citenamefont {Dubrovinsky}, \citenamefont {Braun}, \citenamefont {Celliers}, \citenamefont {Collins}, \citenamefont {Eggert},\ and\ \citenamefont {Jeanloz}}]{millot2015}%
  \BibitemOpen
  \bibfield  {author} {\bibinfo {author} {\bibfnamefont {M.}~\bibnamefont {Millot}}, \bibinfo {author} {\bibfnamefont {N.}~\bibnamefont {Dubrovinskaia}}, \bibinfo {author} {\bibfnamefont {A.}~\bibnamefont {Černok}}, \bibinfo {author} {\bibfnamefont {S.}~\bibnamefont {Blaha}}, \bibinfo {author} {\bibfnamefont {L.}~\bibnamefont {Dubrovinsky}}, \bibinfo {author} {\bibfnamefont {D.~G.}\ \bibnamefont {Braun}}, \bibinfo {author} {\bibfnamefont {P.~M.}\ \bibnamefont {Celliers}}, \bibinfo {author} {\bibfnamefont {G.~W.}\ \bibnamefont {Collins}}, \bibinfo {author} {\bibfnamefont {J.~H.}\ \bibnamefont {Eggert}},\ and\ \bibinfo {author} {\bibfnamefont {R.}~\bibnamefont {Jeanloz}},\ }\bibfield  {title} {\bibinfo {title} {Shock compression of stishovite and melting of silica at planetary interior conditions},\ }\href {https://doi.org/10.1126/science.1261507} {\bibfield  {journal} {\bibinfo  {journal} {Science}\ }\textbf {\bibinfo {volume} {347}},\ \bibinfo {pages} {418} (\bibinfo {year} {2015})}\BibitemShut {NoStop}%
\bibitem [{\citenamefont {Kohn}\ and\ \citenamefont {Sham}(1965)}]{kohn1965}%
  \BibitemOpen
  \bibfield  {author} {\bibinfo {author} {\bibfnamefont {W.}~\bibnamefont {Kohn}}\ and\ \bibinfo {author} {\bibfnamefont {L.~J.}\ \bibnamefont {Sham}},\ }\bibfield  {title} {\bibinfo {title} {Self-consistent equations including exchange and correlation effects},\ }\href {https://doi.org/10.1103/PhysRev.140.A1133} {\bibfield  {journal} {\bibinfo  {journal} {Phys.\ Rev.}\ }\textbf {\bibinfo {volume} {140}},\ \bibinfo {pages} {A1133} (\bibinfo {year} {1965})}\BibitemShut {NoStop}%
\bibitem [{\citenamefont {Geng}\ and\ \citenamefont {Mohn}(2024)}]{geng2024}%
  \BibitemOpen
  \bibfield  {author} {\bibinfo {author} {\bibfnamefont {M.}~\bibnamefont {Geng}}\ and\ \bibinfo {author} {\bibfnamefont {C.~E.}\ \bibnamefont {Mohn}},\ }\bibfield  {title} {\bibinfo {title} {Ab initio constraints on the melting of silica at high pressures up to 500 gpa},\ }\href {https://doi.org/10.1103/PhysRevB.109.024106} {\bibfield  {journal} {\bibinfo  {journal} {Phys.\ Rev.\ B}\ }\textbf {\bibinfo {volume} {109}},\ \bibinfo {pages} {024106} (\bibinfo {year} {2024})}\BibitemShut {NoStop}%
\bibitem [{\citenamefont {González-Cataldo}\ \emph {et~al.}(2016{\natexlab{a}})\citenamefont {González-Cataldo}, \citenamefont {Davis},\ and\ \citenamefont {Gutiérrez}}]{gonzalez2016}%
  \BibitemOpen
  \bibfield  {author} {\bibinfo {author} {\bibfnamefont {F.}~\bibnamefont {González-Cataldo}}, \bibinfo {author} {\bibfnamefont {S.}~\bibnamefont {Davis}},\ and\ \bibinfo {author} {\bibfnamefont {G.}~\bibnamefont {Gutiérrez}},\ }\bibfield  {title} {\bibinfo {title} {Melting curve of sio$_2$ at multimegabar pressures: Implications for gas giants and super-earths},\ }\href {https://doi.org/10.1038/srep26537} {\bibfield  {journal} {\bibinfo  {journal} {Sci.\ Rep.}\ }\textbf {\bibinfo {volume} {6}},\ \bibinfo {pages} {26537} (\bibinfo {year} {2016}{\natexlab{a}})}\BibitemShut {NoStop}%
\bibitem [{\citenamefont {González-Cataldo}\ \emph {et~al.}(2016{\natexlab{b}})\citenamefont {González-Cataldo}, \citenamefont {Davis},\ and\ \citenamefont {Gutiérrez}}]{gonzalez2016conf}%
  \BibitemOpen
  \bibfield  {author} {\bibinfo {author} {\bibfnamefont {F.}~\bibnamefont {González-Cataldo}}, \bibinfo {author} {\bibfnamefont {S.}~\bibnamefont {Davis}},\ and\ \bibinfo {author} {\bibfnamefont {G.}~\bibnamefont {Gutiérrez}},\ }\bibfield  {title} {\bibinfo {title} {Z method calculations to determine the melting curve of silica at high pressures},\ }\href {https://doi.org/10.1088/1742-6596/720/1/012032} {\bibfield  {journal} {\bibinfo  {journal} {J.\ Phys.:\ Conf.\ Ser.}\ }\textbf {\bibinfo {volume} {720}},\ \bibinfo {pages} {012032} (\bibinfo {year} {2016}{\natexlab{b}})}\BibitemShut {NoStop}%
\bibitem [{\citenamefont {Das}\ \emph {et~al.}(2020)\citenamefont {Das}, \citenamefont {Mohn}, \citenamefont {Brodholt},\ and\ \citenamefont {Trønnes}}]{das2020}%
  \BibitemOpen
  \bibfield  {author} {\bibinfo {author} {\bibfnamefont {P.~K.}\ \bibnamefont {Das}}, \bibinfo {author} {\bibfnamefont {C.~E.}\ \bibnamefont {Mohn}}, \bibinfo {author} {\bibfnamefont {J.~P.}\ \bibnamefont {Brodholt}},\ and\ \bibinfo {author} {\bibfnamefont {R.~G.}\ \bibnamefont {Trønnes}},\ }\bibfield  {title} {\bibinfo {title} {High-pressure silica phase transitions: Implications for deep mantle dynamics and silica crystallization in the protocore},\ }\href {https://doi.org/10.2138/am-2020-7299} {\bibfield  {journal} {\bibinfo  {journal} {Am.\ Mineral.}\ }\textbf {\bibinfo {volume} {105}},\ \bibinfo {pages} {1014} (\bibinfo {year} {2020})}\BibitemShut {NoStop}%
\bibitem [{\citenamefont {Ceperley}\ and\ \citenamefont {Alder}(1980)}]{ceperley1980}%
  \BibitemOpen
  \bibfield  {author} {\bibinfo {author} {\bibfnamefont {D.~M.}\ \bibnamefont {Ceperley}}\ and\ \bibinfo {author} {\bibfnamefont {B.~J.}\ \bibnamefont {Alder}},\ }\bibfield  {title} {\bibinfo {title} {Ground state of the electron gas by a stochastic method},\ }\href {https://doi.org/10.1103/PhysRevLett.45.566} {\bibfield  {journal} {\bibinfo  {journal} {Phys.\ Rev.\ Lett.}\ }\textbf {\bibinfo {volume} {45}},\ \bibinfo {pages} {566} (\bibinfo {year} {1980})}\BibitemShut {NoStop}%
\bibitem [{\citenamefont {Perdew}\ and\ \citenamefont {Zunger}(1981)}]{perdew1981}%
  \BibitemOpen
  \bibfield  {author} {\bibinfo {author} {\bibfnamefont {J.~P.}\ \bibnamefont {Perdew}}\ and\ \bibinfo {author} {\bibfnamefont {A.}~\bibnamefont {Zunger}},\ }\bibfield  {title} {\bibinfo {title} {Self-interaction correction to density-functional approximations for many-electron systems},\ }\href {https://doi.org/10.1103/PhysRevB.23.5048} {\bibfield  {journal} {\bibinfo  {journal} {Phys.\ Rev.\ B}\ }\textbf {\bibinfo {volume} {23}},\ \bibinfo {pages} {5048} (\bibinfo {year} {1981})}\BibitemShut {NoStop}%
\bibitem [{\citenamefont {Lee}\ \emph {et~al.}(1988)\citenamefont {Lee}, \citenamefont {Yang},\ and\ \citenamefont {Parr}}]{lee1988}%
  \BibitemOpen
  \bibfield  {author} {\bibinfo {author} {\bibfnamefont {C.}~\bibnamefont {Lee}}, \bibinfo {author} {\bibfnamefont {W.}~\bibnamefont {Yang}},\ and\ \bibinfo {author} {\bibfnamefont {R.~G.}\ \bibnamefont {Parr}},\ }\bibfield  {title} {\bibinfo {title} {Development of the colle-salvetti correlation-energy formula into a functional of the electron density},\ }\href {https://doi.org/10.1103/PhysRevB.37.785} {\bibfield  {journal} {\bibinfo  {journal} {Phys.\ Rev.\ B}\ }\textbf {\bibinfo {volume} {37}},\ \bibinfo {pages} {785} (\bibinfo {year} {1988})}\BibitemShut {NoStop}%
\bibitem [{\citenamefont {Becke}(1988)}]{becke1988}%
  \BibitemOpen
  \bibfield  {author} {\bibinfo {author} {\bibfnamefont {A.~D.}\ \bibnamefont {Becke}},\ }\bibfield  {title} {\bibinfo {title} {Density-functional exchange-energy approximation with correct asymptotic behavior},\ }\href {https://doi.org/10.1103/PhysRevA.38.3098} {\bibfield  {journal} {\bibinfo  {journal} {Phys.\ Rev.\ A}\ }\textbf {\bibinfo {volume} {38}},\ \bibinfo {pages} {3098} (\bibinfo {year} {1988})}\BibitemShut {NoStop}%
\bibitem [{\citenamefont {Perdew}\ \emph {et~al.}(1996)\citenamefont {Perdew}, \citenamefont {Burke},\ and\ \citenamefont {Ernzerhof}}]{perdew1996}%
  \BibitemOpen
  \bibfield  {author} {\bibinfo {author} {\bibfnamefont {J.~P.}\ \bibnamefont {Perdew}}, \bibinfo {author} {\bibfnamefont {K.}~\bibnamefont {Burke}},\ and\ \bibinfo {author} {\bibfnamefont {M.}~\bibnamefont {Ernzerhof}},\ }\bibfield  {title} {\bibinfo {title} {Generalized gradient approximation made simple},\ }\href {https://doi.org/10.1103/PhysRevLett.77.3865} {\bibfield  {journal} {\bibinfo  {journal} {Phys.\ Rev.\ Lett.}\ }\textbf {\bibinfo {volume} {77}},\ \bibinfo {pages} {3865} (\bibinfo {year} {1996})}\BibitemShut {NoStop}%
\bibitem [{\citenamefont {Perdew}\ and\ \citenamefont {Schmidt}(2001)}]{perdew2001}%
  \BibitemOpen
  \bibfield  {author} {\bibinfo {author} {\bibfnamefont {J.~P.}\ \bibnamefont {Perdew}}\ and\ \bibinfo {author} {\bibfnamefont {K.}~\bibnamefont {Schmidt}},\ }\bibfield  {title} {\bibinfo {title} {Jacob’s ladder of density functional approximations for the exchange-correlation energy},\ }in\ \href@noop {} {\emph {\bibinfo {booktitle} {AIP Conference Proceedings}}},\ Vol.\ \bibinfo {volume} {577},\ \bibinfo {editor} {edited by\ \bibinfo {editor} {\bibfnamefont {V.~V.}\ \bibnamefont {Doren}}, \bibinfo {editor} {\bibfnamefont {C.~V.}\ \bibnamefont {Alsenoy}},\ and\ \bibinfo {editor} {\bibfnamefont {P.}~\bibnamefont {Geerlings}}}\ (\bibinfo  {publisher} {American Institute of Physics},\ \bibinfo {address} {Melville, NY},\ \bibinfo {year} {2001})\ pp.\ \bibinfo {pages} {1--20}\BibitemShut {NoStop}%
\bibitem [{\citenamefont {Sun}\ \emph {et~al.}(2015)\citenamefont {Sun}, \citenamefont {Ruzsinszky},\ and\ \citenamefont {Perdew}}]{sun2015}%
  \BibitemOpen
  \bibfield  {author} {\bibinfo {author} {\bibfnamefont {J.}~\bibnamefont {Sun}}, \bibinfo {author} {\bibfnamefont {A.}~\bibnamefont {Ruzsinszky}},\ and\ \bibinfo {author} {\bibfnamefont {J.~P.}\ \bibnamefont {Perdew}},\ }\bibfield  {title} {\bibinfo {title} {Strongly constrained and appropriately normed semilocal density functional},\ }\href {https://doi.org/10.1103/PhysRevLett.115.036402} {\bibfield  {journal} {\bibinfo  {journal} {Phys.\ Rev.\ Lett.}\ }\textbf {\bibinfo {volume} {115}},\ \bibinfo {pages} {036402} (\bibinfo {year} {2015})}\BibitemShut {NoStop}%
\bibitem [{\citenamefont {Sun}\ \emph {et~al.}(2016)\citenamefont {Sun}, \citenamefont {Remsing}, \citenamefont {Zhang}, \citenamefont {Sun}, \citenamefont {Ruzsinszky}, \citenamefont {Peng}, \citenamefont {Yang}, \citenamefont {Paul}, \citenamefont {Waghmare}, \citenamefont {Wu}, \citenamefont {Klein},\ and\ \citenamefont {Perdew}}]{sun2016}%
  \BibitemOpen
  \bibfield  {author} {\bibinfo {author} {\bibfnamefont {J.}~\bibnamefont {Sun}}, \bibinfo {author} {\bibfnamefont {R.~C.}\ \bibnamefont {Remsing}}, \bibinfo {author} {\bibfnamefont {Y.}~\bibnamefont {Zhang}}, \bibinfo {author} {\bibfnamefont {Z.}~\bibnamefont {Sun}}, \bibinfo {author} {\bibfnamefont {A.}~\bibnamefont {Ruzsinszky}}, \bibinfo {author} {\bibfnamefont {H.}~\bibnamefont {Peng}}, \bibinfo {author} {\bibfnamefont {Z.}~\bibnamefont {Yang}}, \bibinfo {author} {\bibfnamefont {A.}~\bibnamefont {Paul}}, \bibinfo {author} {\bibfnamefont {U.}~\bibnamefont {Waghmare}}, \bibinfo {author} {\bibfnamefont {X.}~\bibnamefont {Wu}}, \bibinfo {author} {\bibfnamefont {M.~L.}\ \bibnamefont {Klein}},\ and\ \bibinfo {author} {\bibfnamefont {J.~P.}\ \bibnamefont {Perdew}},\ }\bibfield  {title} {\bibinfo {title} {Accurate first-principles structures and energies of diversely bonded systems from an efficient density functional},\ }\href {https://doi.org/10.1038/nchem.2535} {\bibfield  {journal} {\bibinfo  {journal} {Nature Chem.}\ }\textbf {\bibinfo {volume} {8}},\ \bibinfo {pages} {831} (\bibinfo {year} {2016})}\BibitemShut {NoStop}%
\bibitem [{\citenamefont {Ning}\ \emph {et~al.}(2022)\citenamefont {Ning}, \citenamefont {Furness},\ and\ \citenamefont {Sun}}]{ning2022}%
  \BibitemOpen
  \bibfield  {author} {\bibinfo {author} {\bibfnamefont {J.}~\bibnamefont {Ning}}, \bibinfo {author} {\bibfnamefont {J.~W.}\ \bibnamefont {Furness}},\ and\ \bibinfo {author} {\bibfnamefont {J.}~\bibnamefont {Sun}},\ }\bibfield  {title} {\bibinfo {title} {Reliable lattice dynamics from an efficient density functional approximation},\ }\href {https://doi.org/10.1021/acs.chemmater.1c03222} {\bibfield  {journal} {\bibinfo  {journal} {Chem.\ Mater.}\ }\textbf {\bibinfo {volume} {34}},\ \bibinfo {pages} {2562} (\bibinfo {year} {2022})}\BibitemShut {NoStop}%
\bibitem [{\citenamefont {Rang}\ and\ \citenamefont {Kresse}(2019)}]{rang2019}%
  \BibitemOpen
  \bibfield  {author} {\bibinfo {author} {\bibfnamefont {M.}~\bibnamefont {Rang}}\ and\ \bibinfo {author} {\bibfnamefont {G.}~\bibnamefont {Kresse}},\ }\bibfield  {title} {\bibinfo {title} {First-principles study of the melting temperature of mgo},\ }\href {https://doi.org/10.1103/PhysRevB.99.184103} {\bibfield  {journal} {\bibinfo  {journal} {Phys.\ Rev.\ B}\ }\textbf {\bibinfo {volume} {99}},\ \bibinfo {pages} {184103} (\bibinfo {year} {2019})}\BibitemShut {NoStop}%
\bibitem [{\citenamefont {Jinnouchi}\ \emph {et~al.}(2019)\citenamefont {Jinnouchi}, \citenamefont {Karsai},\ and\ \citenamefont {Kresse}}]{jinnouchi2019}%
  \BibitemOpen
  \bibfield  {author} {\bibinfo {author} {\bibfnamefont {R.}~\bibnamefont {Jinnouchi}}, \bibinfo {author} {\bibfnamefont {F.}~\bibnamefont {Karsai}},\ and\ \bibinfo {author} {\bibfnamefont {G.}~\bibnamefont {Kresse}},\ }\bibfield  {title} {\bibinfo {title} {On-the-fly machine learning force field generation: Application to melting points},\ }\href {https://doi.org/10.1103/PhysRevB.100.014105} {\bibfield  {journal} {\bibinfo  {journal} {Phys.\ Rev.\ B}\ }\textbf {\bibinfo {volume} {100}},\ \bibinfo {pages} {014105} (\bibinfo {year} {2019})}\BibitemShut {NoStop}%
\bibitem [{\citenamefont {Behler}\ and\ \citenamefont {Parrinello}(2007)}]{behler2007}%
  \BibitemOpen
  \bibfield  {author} {\bibinfo {author} {\bibfnamefont {J.}~\bibnamefont {Behler}}\ and\ \bibinfo {author} {\bibfnamefont {M.}~\bibnamefont {Parrinello}},\ }\bibfield  {title} {\bibinfo {title} {Generalized neural-network representation of high-dimensional potential-energy surfaces},\ }\href {https://doi.org/10.1103/PhysRevLett.98.146401} {\bibfield  {journal} {\bibinfo  {journal} {Phys.\ Rev.\ Lett.}\ }\textbf {\bibinfo {volume} {98}},\ \bibinfo {pages} {146401} (\bibinfo {year} {2007})}\BibitemShut {NoStop}%
\bibitem [{\citenamefont {Zhang}\ \emph {et~al.}(2018)\citenamefont {Zhang}, \citenamefont {Han}, \citenamefont {Wang}, \citenamefont {Car},\ and\ \citenamefont {E}}]{zhang2018}%
  \BibitemOpen
  \bibfield  {author} {\bibinfo {author} {\bibfnamefont {L.}~\bibnamefont {Zhang}}, \bibinfo {author} {\bibfnamefont {J.}~\bibnamefont {Han}}, \bibinfo {author} {\bibfnamefont {H.}~\bibnamefont {Wang}}, \bibinfo {author} {\bibfnamefont {R.}~\bibnamefont {Car}},\ and\ \bibinfo {author} {\bibfnamefont {W.}~\bibnamefont {E}},\ }\bibfield  {title} {\bibinfo {title} {Deep potential molecular dynamics: A scalable model with the accuracy of quantum mechanics},\ }\href {https://doi.org/10.1103/PhysRevLett.120.143001} {\bibfield  {journal} {\bibinfo  {journal} {Phys.\ Rev.\ Lett.}\ }\textbf {\bibinfo {volume} {120}},\ \bibinfo {pages} {143001} (\bibinfo {year} {2018})}\BibitemShut {NoStop}%
\bibitem [{\citenamefont {Yang}\ \emph {et~al.}(2021)\citenamefont {Yang}, \citenamefont {Karmakar},\ and\ \citenamefont {Parrinello}}]{yang2021}%
  \BibitemOpen
  \bibfield  {author} {\bibinfo {author} {\bibfnamefont {M.}~\bibnamefont {Yang}}, \bibinfo {author} {\bibfnamefont {T.}~\bibnamefont {Karmakar}},\ and\ \bibinfo {author} {\bibfnamefont {M.}~\bibnamefont {Parrinello}},\ }\bibfield  {title} {\bibinfo {title} {Liquid-liquid critical point in phosphorus},\ }\href {https://doi.org/10.1103/PhysRevLett.127.080603} {\bibfield  {journal} {\bibinfo  {journal} {Phys.\ Rev.\ Lett.}\ }\textbf {\bibinfo {volume} {127}},\ \bibinfo {pages} {080603} (\bibinfo {year} {2021})}\BibitemShut {NoStop}%
\bibitem [{\citenamefont {Deng}\ \emph {et~al.}(2023)\citenamefont {Deng}, \citenamefont {Niu}, \citenamefont {Hu}, \citenamefont {Chen},\ and\ \citenamefont {Stixrude}}]{deng2023}%
  \BibitemOpen
  \bibfield  {author} {\bibinfo {author} {\bibfnamefont {J.}~\bibnamefont {Deng}}, \bibinfo {author} {\bibfnamefont {H.}~\bibnamefont {Niu}}, \bibinfo {author} {\bibfnamefont {J.}~\bibnamefont {Hu}}, \bibinfo {author} {\bibfnamefont {M.}~\bibnamefont {Chen}},\ and\ \bibinfo {author} {\bibfnamefont {L.}~\bibnamefont {Stixrude}},\ }\bibfield  {title} {\bibinfo {title} {Melting of mgsio$_3$ determined by machine learning potentials},\ }\href {https://doi.org/10.1103/PhysRevB.107.064103} {\bibfield  {journal} {\bibinfo  {journal} {Phys.\ Rev.\ B}\ }\textbf {\bibinfo {volume} {107}},\ \bibinfo {pages} {064103} (\bibinfo {year} {2023})}\BibitemShut {NoStop}%
\bibitem [{\citenamefont {Perdew}\ \emph {et~al.}(2008)\citenamefont {Perdew}, \citenamefont {Ruzsinszky}, \citenamefont {Csonka}, \citenamefont {Vydrov}, \citenamefont {Scuseria}, \citenamefont {Constantin}, \citenamefont {Zhou},\ and\ \citenamefont {Burke}}]{perdew2008}%
  \BibitemOpen
  \bibfield  {author} {\bibinfo {author} {\bibfnamefont {J.~P.}\ \bibnamefont {Perdew}}, \bibinfo {author} {\bibfnamefont {A.}~\bibnamefont {Ruzsinszky}}, \bibinfo {author} {\bibfnamefont {G.~I.}\ \bibnamefont {Csonka}}, \bibinfo {author} {\bibfnamefont {O.~A.}\ \bibnamefont {Vydrov}}, \bibinfo {author} {\bibfnamefont {G.~E.}\ \bibnamefont {Scuseria}}, \bibinfo {author} {\bibfnamefont {L.~A.}\ \bibnamefont {Constantin}}, \bibinfo {author} {\bibfnamefont {X.}~\bibnamefont {Zhou}},\ and\ \bibinfo {author} {\bibfnamefont {K.}~\bibnamefont {Burke}},\ }\bibfield  {title} {\bibinfo {title} {Restoring the density-gradient expansion for exchange in solids and surfaces},\ }\href {https://doi.org/10.1103/PhysRevLett.100.136406} {\bibfield  {journal} {\bibinfo  {journal} {Phys.\ Rev.\ Lett.}\ }\textbf {\bibinfo {volume} {100}},\ \bibinfo {pages} {136406} (\bibinfo {year} {2008})}\BibitemShut {NoStop}%
\bibitem [{\citenamefont {Valsson}\ and\ \citenamefont {Parrinello}(2014)}]{valsson2014}%
  \BibitemOpen
  \bibfield  {author} {\bibinfo {author} {\bibfnamefont {O.}~\bibnamefont {Valsson}}\ and\ \bibinfo {author} {\bibfnamefont {M.}~\bibnamefont {Parrinello}},\ }\bibfield  {title} {\bibinfo {title} {Variational approach to enhanced sampling and free energy calculations},\ }\href {https://doi.org/10.1103/PhysRevLett.113.090601} {\bibfield  {journal} {\bibinfo  {journal} {Phys.\ Rev.\ Lett.}\ }\textbf {\bibinfo {volume} {113}},\ \bibinfo {pages} {090601} (\bibinfo {year} {2014})}\BibitemShut {NoStop}%
\bibitem [{\citenamefont {Niu}\ \emph {et~al.}(2018)\citenamefont {Niu}, \citenamefont {Piaggi}, \citenamefont {Invernizzi},\ and\ \citenamefont {Parrinello}}]{niu2018}%
  \BibitemOpen
  \bibfield  {author} {\bibinfo {author} {\bibfnamefont {H.}~\bibnamefont {Niu}}, \bibinfo {author} {\bibfnamefont {P.~M.}\ \bibnamefont {Piaggi}}, \bibinfo {author} {\bibfnamefont {M.}~\bibnamefont {Invernizzi}},\ and\ \bibinfo {author} {\bibfnamefont {M.}~\bibnamefont {Parrinello}},\ }\bibfield  {title} {\bibinfo {title} {Molecular dynamics simulations of liquid silica crystallization},\ }\href {https://doi.org/10.1073/pnas.1803919115} {\bibfield  {journal} {\bibinfo  {journal} {Proc.\ Natl.\ Acad.\ Sci.\ U.S.A.}\ }\textbf {\bibinfo {volume} {115}},\ \bibinfo {pages} {5348} (\bibinfo {year} {2018})}\BibitemShut {NoStop}%
\bibitem [{\citenamefont {Plimpton}(1995)}]{plimpton1995}%
  \BibitemOpen
  \bibfield  {author} {\bibinfo {author} {\bibfnamefont {S.}~\bibnamefont {Plimpton}},\ }\bibfield  {title} {\bibinfo {title} {Fast parallel algorithms for short-range molecular dynamics},\ }\href {https://doi.org/10.1006/jcph.1995.1039} {\bibfield  {journal} {\bibinfo  {journal} {J.\ Comput.\ Phys.}\ }\textbf {\bibinfo {volume} {117}},\ \bibinfo {pages} {1} (\bibinfo {year} {1995})}\BibitemShut {NoStop}%
\bibitem [{\citenamefont {Thompson}\ \emph {et~al.}(2022)\citenamefont {Thompson}, \citenamefont {Aktulga}, \citenamefont {Berger}, \citenamefont {Bolintineanu}, \citenamefont {Brown}, \citenamefont {Crozier}, \citenamefont {in~'t Veld}, \citenamefont {Kohlmeyer}, \citenamefont {Moore}, \citenamefont {Nguyen}, \citenamefont {Shan}, \citenamefont {Stevens}, \citenamefont {Tranchida}, \citenamefont {Trott},\ and\ \citenamefont {Plimpton}}]{thompson2022}%
  \BibitemOpen
  \bibfield  {author} {\bibinfo {author} {\bibfnamefont {A.~P.}\ \bibnamefont {Thompson}}, \bibinfo {author} {\bibfnamefont {H.~M.}\ \bibnamefont {Aktulga}}, \bibinfo {author} {\bibfnamefont {R.}~\bibnamefont {Berger}}, \bibinfo {author} {\bibfnamefont {D.~S.}\ \bibnamefont {Bolintineanu}}, \bibinfo {author} {\bibfnamefont {W.~M.}\ \bibnamefont {Brown}}, \bibinfo {author} {\bibfnamefont {P.~S.}\ \bibnamefont {Crozier}}, \bibinfo {author} {\bibfnamefont {P.~J.}\ \bibnamefont {in~'t Veld}}, \bibinfo {author} {\bibfnamefont {A.}~\bibnamefont {Kohlmeyer}}, \bibinfo {author} {\bibfnamefont {S.~G.}\ \bibnamefont {Moore}}, \bibinfo {author} {\bibfnamefont {T.~D.}\ \bibnamefont {Nguyen}}, \bibinfo {author} {\bibfnamefont {R.}~\bibnamefont {Shan}}, \bibinfo {author} {\bibfnamefont {M.~J.}\ \bibnamefont {Stevens}}, \bibinfo {author} {\bibfnamefont {J.}~\bibnamefont {Tranchida}}, \bibinfo {author} {\bibfnamefont {C.}~\bibnamefont {Trott}},\ and\ \bibinfo {author} {\bibfnamefont {S.~J.}\ \bibnamefont {Plimpton}},\ }\bibfield  {title} {\bibinfo {title} {Lammps -- a flexible simulation tool for particle-based materials modeling at the atomic, meso, and continuum scales},\ }\href {https://doi.org/10.1016/j.cpc.2021.108171} {\bibfield  {journal} {\bibinfo  {journal} {Comput.\ Phys.\ Commun.}\ }\textbf {\bibinfo {volume} {271}},\ \bibinfo {pages} {108171} (\bibinfo {year} {2022})}\BibitemShut {NoStop}%
\bibitem [{\citenamefont {Zeng}\ \emph {et~al.}(2025)\citenamefont {Zeng}, \citenamefont {Zhang}, \citenamefont {Peng}, \citenamefont {Zhang}, \citenamefont {He}, \citenamefont {Wang}, \citenamefont {Liu}, \citenamefont {Bi}, \citenamefont {Li}, \citenamefont {Cai}, \citenamefont {Zhang}, \citenamefont {Du}, \citenamefont {Zhu}, \citenamefont {Mo}, \citenamefont {Huang}, \citenamefont {Zeng}, \citenamefont {Shi}, \citenamefont {Qin}, \citenamefont {Yu}, \citenamefont {Wang} \emph {et~al.}}]{zeng2025}%
  \BibitemOpen
  \bibfield  {author} {\bibinfo {author} {\bibfnamefont {J.}~\bibnamefont {Zeng}}, \bibinfo {author} {\bibfnamefont {D.}~\bibnamefont {Zhang}}, \bibinfo {author} {\bibfnamefont {A.}~\bibnamefont {Peng}}, \bibinfo {author} {\bibfnamefont {X.}~\bibnamefont {Zhang}}, \bibinfo {author} {\bibfnamefont {S.}~\bibnamefont {He}}, \bibinfo {author} {\bibfnamefont {Y.}~\bibnamefont {Wang}}, \bibinfo {author} {\bibfnamefont {X.}~\bibnamefont {Liu}}, \bibinfo {author} {\bibfnamefont {H.}~\bibnamefont {Bi}}, \bibinfo {author} {\bibfnamefont {Y.}~\bibnamefont {Li}}, \bibinfo {author} {\bibfnamefont {C.}~\bibnamefont {Cai}}, \bibinfo {author} {\bibfnamefont {C.}~\bibnamefont {Zhang}}, \bibinfo {author} {\bibfnamefont {Y.}~\bibnamefont {Du}}, \bibinfo {author} {\bibfnamefont {J.-X.}\ \bibnamefont {Zhu}}, \bibinfo {author} {\bibfnamefont {P.}~\bibnamefont {Mo}}, \bibinfo {author} {\bibfnamefont {Z.}~\bibnamefont {Huang}}, \bibinfo {author} {\bibfnamefont {Q.}~\bibnamefont {Zeng}}, \bibinfo {author} {\bibfnamefont {S.}~\bibnamefont {Shi}}, \bibinfo {author} {\bibfnamefont {X.}~\bibnamefont {Qin}}, \bibinfo {author} {\bibfnamefont {Z.}~\bibnamefont {Yu}}, \bibinfo {author} {\bibfnamefont {H.}~\bibnamefont {Wang}}, \emph {et~al.},\ }\bibfield  {title} {\bibinfo {title} {Deepmd-kit v3: A multiple-backend framework for machine learning potentials},\ }\href {https://doi.org/10.1021/acs.jctc.5c00340} {\bibfield  {journal} {\bibinfo  {journal} {J.\ Chem.\ Theory Comput.}\ }\textbf {\bibinfo {volume} {21}},\ \bibinfo {pages} {4375} (\bibinfo {year} {2025})}\BibitemShut {NoStop}%
\bibitem [{\citenamefont {Tribello}\ \emph {et~al.}(2014)\citenamefont {Tribello}, \citenamefont {Bonomi}, \citenamefont {Branduardi}, \citenamefont {Camilloni},\ and\ \citenamefont {Bussi}}]{tribello2014}%
  \BibitemOpen
  \bibfield  {author} {\bibinfo {author} {\bibfnamefont {G.~A.}\ \bibnamefont {Tribello}}, \bibinfo {author} {\bibfnamefont {M.}~\bibnamefont {Bonomi}}, \bibinfo {author} {\bibfnamefont {D.}~\bibnamefont {Branduardi}}, \bibinfo {author} {\bibfnamefont {C.}~\bibnamefont {Camilloni}},\ and\ \bibinfo {author} {\bibfnamefont {G.}~\bibnamefont {Bussi}},\ }\bibfield  {title} {\bibinfo {title} {Plumed 2: New feathers for an old bird},\ }\href {https://doi.org/10.1016/j.cpc.2013.09.018} {\bibfield  {journal} {\bibinfo  {journal} {Comput.\ Phys.\ Commun.}\ }\textbf {\bibinfo {volume} {185}},\ \bibinfo {pages} {604} (\bibinfo {year} {2014})}\BibitemShut {NoStop}%
\bibitem [{\citenamefont {Kresse}\ and\ \citenamefont {Furthmüller}(1996)}]{kresse1996}%
  \BibitemOpen
  \bibfield  {author} {\bibinfo {author} {\bibfnamefont {G.}~\bibnamefont {Kresse}}\ and\ \bibinfo {author} {\bibfnamefont {J.}~\bibnamefont {Furthmüller}},\ }\bibfield  {title} {\bibinfo {title} {Efficiency of ab-initio total energy calculations for metals and semiconductors using a plane-wave basis set},\ }\href {https://doi.org/10.1016/0927-0256(96)00008-0} {\bibfield  {journal} {\bibinfo  {journal} {Comput.\ Mater.\ Sci.}\ }\textbf {\bibinfo {volume} {6}},\ \bibinfo {pages} {15} (\bibinfo {year} {1996})}\BibitemShut {NoStop}%
\bibitem [{\citenamefont {Kresse}\ and\ \citenamefont {Joubert}(1999)}]{kresse1999}%
  \BibitemOpen
  \bibfield  {author} {\bibinfo {author} {\bibfnamefont {G.}~\bibnamefont {Kresse}}\ and\ \bibinfo {author} {\bibfnamefont {D.}~\bibnamefont {Joubert}},\ }\bibfield  {title} {\bibinfo {title} {From ultrasoft pseudopotentials to the projector augmented-wave method},\ }\href {https://doi.org/10.1103/PhysRevB.59.1758} {\bibfield  {journal} {\bibinfo  {journal} {Phys.\ Rev.\ B}\ }\textbf {\bibinfo {volume} {59}},\ \bibinfo {pages} {1758} (\bibinfo {year} {1999})}\BibitemShut {NoStop}%
\bibitem [{\citenamefont {Deng}\ and\ \citenamefont {Stixrude}(2021)}]{deng2021}%
  \BibitemOpen
  \bibfield  {author} {\bibinfo {author} {\bibfnamefont {J.}~\bibnamefont {Deng}}\ and\ \bibinfo {author} {\bibfnamefont {L.}~\bibnamefont {Stixrude}},\ }\bibfield  {title} {\bibinfo {title} {Thermal conductivity of silicate liquid determined by machine learning potentials},\ }\href {https://doi.org/10.1029/2021GL093806} {\bibfield  {journal} {\bibinfo  {journal} {Geophys.\ Res.\ Lett.}\ }\textbf {\bibinfo {volume} {48}},\ \bibinfo {pages} {e2021GL093806} (\bibinfo {year} {2021})}\BibitemShut {NoStop}%
\bibitem [{\citenamefont {Deng}\ \emph {et~al.}(2017)\citenamefont {Deng}, \citenamefont {Du}, \citenamefont {Benedetti},\ and\ \citenamefont {Lee}}]{deng2017}%
  \BibitemOpen
  \bibfield  {author} {\bibinfo {author} {\bibfnamefont {J.}~\bibnamefont {Deng}}, \bibinfo {author} {\bibfnamefont {Z.}~\bibnamefont {Du}}, \bibinfo {author} {\bibfnamefont {L.~R.}\ \bibnamefont {Benedetti}},\ and\ \bibinfo {author} {\bibfnamefont {K.~K.~M.}\ \bibnamefont {Lee}},\ }\bibfield  {title} {\bibinfo {title} {The influence of wavelength-dependent absorption and temperature gradients on temperature determination in laser-heated diamond-anvil cells},\ }\href {https://doi.org/10.1063/1.4973344} {\bibfield  {journal} {\bibinfo  {journal} {J.\ Appl.\ Phys.}\ }\textbf {\bibinfo {volume} {121}},\ \bibinfo {pages} {025901} (\bibinfo {year} {2017})}\BibitemShut {NoStop}%
\bibitem [{\citenamefont {Bourova}\ and\ \citenamefont {Richet}(1998)}]{bourova1998}%
  \BibitemOpen
  \bibfield  {author} {\bibinfo {author} {\bibfnamefont {E.}~\bibnamefont {Bourova}}\ and\ \bibinfo {author} {\bibfnamefont {P.}~\bibnamefont {Richet}},\ }\bibfield  {title} {\bibinfo {title} {Quartz and cristobalite: High-temperature cell parameters and volumes of fusion},\ }\href {https://doi.org/10.1029/98GL01581} {\bibfield  {journal} {\bibinfo  {journal} {Geophys.\ Res.\ Lett.}\ }\textbf {\bibinfo {volume} {25}},\ \bibinfo {pages} {2333} (\bibinfo {year} {1998})}\BibitemShut {NoStop}%
\bibitem [{\citenamefont {Young}(1991)}]{young1991}%
  \BibitemOpen
  \bibfield  {author} {\bibinfo {author} {\bibfnamefont {D.~A.}\ \bibnamefont {Young}},\ }\href@noop {} {\emph {\bibinfo {title} {Phase Diagrams of the Elements}}}\ (\bibinfo  {publisher} {University of California Press},\ \bibinfo {address} {Berkeley, CA},\ \bibinfo {year} {1991})\BibitemShut {NoStop}%
\bibitem [{\citenamefont {Stixrude}\ and\ \citenamefont {Lithgow-Bertelloni}(2024)}]{stixrude2024}%
  \BibitemOpen
  \bibfield  {author} {\bibinfo {author} {\bibfnamefont {L.}~\bibnamefont {Stixrude}}\ and\ \bibinfo {author} {\bibfnamefont {C.}~\bibnamefont {Lithgow-Bertelloni}},\ }\bibfield  {title} {\bibinfo {title} {Thermodynamics of mantle minerals--iii: the role of iron},\ }\href {https://doi.org/10.1093/gji/ggae126} {\bibfield  {journal} {\bibinfo  {journal} {Geophys.\ J.\ Int.}\ }\textbf {\bibinfo {volume} {237}},\ \bibinfo {pages} {1699} (\bibinfo {year} {2024})}\BibitemShut {NoStop}%
\bibitem [{\citenamefont {Knittle}\ and\ \citenamefont {Jeanloz}(1989)}]{knittle1989}%
  \BibitemOpen
  \bibfield  {author} {\bibinfo {author} {\bibfnamefont {E.}~\bibnamefont {Knittle}}\ and\ \bibinfo {author} {\bibfnamefont {R.}~\bibnamefont {Jeanloz}},\ }\bibfield  {title} {\bibinfo {title} {Melting curve of (mg,fe)sio$_3$ perovskite to 96 gpa: Evidence for a structural transition in lower mantle melts},\ }\href {https://doi.org/10.1029/GL016i005p00421} {\bibfield  {journal} {\bibinfo  {journal} {Geophys.\ Res.\ Lett.}\ }\textbf {\bibinfo {volume} {16}},\ \bibinfo {pages} {421} (\bibinfo {year} {1989})}\BibitemShut {NoStop}%
\bibitem [{\citenamefont {Tsuchiya}\ \emph {et~al.}(2004)\citenamefont {Tsuchiya}, \citenamefont {Caracas},\ and\ \citenamefont {Tsuchiya}}]{tsuchiya2004}%
  \BibitemOpen
  \bibfield  {author} {\bibinfo {author} {\bibfnamefont {T.}~\bibnamefont {Tsuchiya}}, \bibinfo {author} {\bibfnamefont {R.}~\bibnamefont {Caracas}},\ and\ \bibinfo {author} {\bibfnamefont {J.}~\bibnamefont {Tsuchiya}},\ }\bibfield  {title} {\bibinfo {title} {First principles determination of the phase boundaries of high-pressure polymorphs of silica},\ }\href {https://doi.org/10.1029/2004GL019649} {\bibfield  {journal} {\bibinfo  {journal} {Geophys.\ Res.\ Lett.}\ }\textbf {\bibinfo {volume} {31}},\ \bibinfo {pages} {L11610} (\bibinfo {year} {2004})}\BibitemShut {NoStop}%
\bibitem [{\citenamefont {Holmes}\ \emph {et~al.}(1989)\citenamefont {Holmes}, \citenamefont {Moriarty}, \citenamefont {Gathers},\ and\ \citenamefont {Nellis}}]{holmes1989}%
  \BibitemOpen
  \bibfield  {author} {\bibinfo {author} {\bibfnamefont {N.~C.}\ \bibnamefont {Holmes}}, \bibinfo {author} {\bibfnamefont {J.~A.}\ \bibnamefont {Moriarty}}, \bibinfo {author} {\bibfnamefont {G.~R.}\ \bibnamefont {Gathers}},\ and\ \bibinfo {author} {\bibfnamefont {W.~J.}\ \bibnamefont {Nellis}},\ }\bibfield  {title} {\bibinfo {title} {The equation of state of platinum to 660 gpa (6.6 mbar)},\ }\href {https://doi.org/10.1063/1.344047} {\bibfield  {journal} {\bibinfo  {journal} {J.\ Appl.\ Phys.}\ }\textbf {\bibinfo {volume} {66}},\ \bibinfo {pages} {2962} (\bibinfo {year} {1989})}\BibitemShut {NoStop}%
\bibitem [{\citenamefont {Fei}\ \emph {et~al.}(2007)\citenamefont {Fei}, \citenamefont {Ricolleau}, \citenamefont {Frank}, \citenamefont {Mibe}, \citenamefont {Shen},\ and\ \citenamefont {Prakapenka}}]{fei2007}%
  \BibitemOpen
  \bibfield  {author} {\bibinfo {author} {\bibfnamefont {Y.}~\bibnamefont {Fei}}, \bibinfo {author} {\bibfnamefont {A.}~\bibnamefont {Ricolleau}}, \bibinfo {author} {\bibfnamefont {M.}~\bibnamefont {Frank}}, \bibinfo {author} {\bibfnamefont {K.}~\bibnamefont {Mibe}}, \bibinfo {author} {\bibfnamefont {G.}~\bibnamefont {Shen}},\ and\ \bibinfo {author} {\bibfnamefont {V.}~\bibnamefont {Prakapenka}},\ }\bibfield  {title} {\bibinfo {title} {Toward an internally consistent pressure scale},\ }\href {https://doi.org/10.1073/pnas.0609013104} {\bibfield  {journal} {\bibinfo  {journal} {Proc.\ Natl.\ Acad.\ Sci.}\ }\textbf {\bibinfo {volume} {104}},\ \bibinfo {pages} {9182} (\bibinfo {year} {2007})}\BibitemShut {NoStop}%
\bibitem [{\citenamefont {Yokoo}\ \emph {et~al.}(2009)\citenamefont {Yokoo}, \citenamefont {Kawai}, \citenamefont {Nakamura}, \citenamefont {i.~Kondo}, \citenamefont {Tange},\ and\ \citenamefont {Tsuchiya}}]{yokoo2009}%
  \BibitemOpen
  \bibfield  {author} {\bibinfo {author} {\bibfnamefont {M.}~\bibnamefont {Yokoo}}, \bibinfo {author} {\bibfnamefont {N.}~\bibnamefont {Kawai}}, \bibinfo {author} {\bibfnamefont {K.~G.}\ \bibnamefont {Nakamura}}, \bibinfo {author} {\bibfnamefont {K.}~\bibnamefont {i.~Kondo}}, \bibinfo {author} {\bibfnamefont {Y.}~\bibnamefont {Tange}},\ and\ \bibinfo {author} {\bibfnamefont {T.}~\bibnamefont {Tsuchiya}},\ }\bibfield  {title} {\bibinfo {title} {Ultrahigh-pressure scales for gold and platinum at pressures up to 550 gpa},\ }\href {https://doi.org/10.1103/PhysRevB.80.104114} {\bibfield  {journal} {\bibinfo  {journal} {Phys.\ Rev.\ B}\ }\textbf {\bibinfo {volume} {80}},\ \bibinfo {pages} {104114} (\bibinfo {year} {2009})}\BibitemShut {NoStop}%
\bibitem [{\citenamefont {Matsui}\ \emph {et~al.}(2009)\citenamefont {Matsui}, \citenamefont {Ito}, \citenamefont {Katsura}, \citenamefont {Yamazaki}, \citenamefont {Yoshino}, \citenamefont {Yokoyama},\ and\ \citenamefont {Funakoshi}}]{matsui2009}%
  \BibitemOpen
  \bibfield  {author} {\bibinfo {author} {\bibfnamefont {M.}~\bibnamefont {Matsui}}, \bibinfo {author} {\bibfnamefont {E.}~\bibnamefont {Ito}}, \bibinfo {author} {\bibfnamefont {T.}~\bibnamefont {Katsura}}, \bibinfo {author} {\bibfnamefont {D.}~\bibnamefont {Yamazaki}}, \bibinfo {author} {\bibfnamefont {T.}~\bibnamefont {Yoshino}}, \bibinfo {author} {\bibfnamefont {A.}~\bibnamefont {Yokoyama}},\ and\ \bibinfo {author} {\bibfnamefont {K.}~\bibnamefont {Funakoshi}},\ }\bibfield  {title} {\bibinfo {title} {The temperature-pressure-volume equation of state of platinum},\ }\href {https://doi.org/10.1063/1.3054331} {\bibfield  {journal} {\bibinfo  {journal} {J.\ Appl.\ Phys.}\ }\textbf {\bibinfo {volume} {105}},\ \bibinfo {pages} {013505} (\bibinfo {year} {2009})}\BibitemShut {NoStop}%
\bibitem [{\citenamefont {Dorfman}\ \emph {et~al.}(2012)\citenamefont {Dorfman}, \citenamefont {Prakapenka}, \citenamefont {Meng},\ and\ \citenamefont {Duffy}}]{dorfman2012}%
  \BibitemOpen
  \bibfield  {author} {\bibinfo {author} {\bibfnamefont {S.~M.}\ \bibnamefont {Dorfman}}, \bibinfo {author} {\bibfnamefont {V.~B.}\ \bibnamefont {Prakapenka}}, \bibinfo {author} {\bibfnamefont {Y.}~\bibnamefont {Meng}},\ and\ \bibinfo {author} {\bibfnamefont {T.~S.}\ \bibnamefont {Duffy}},\ }\bibfield  {title} {\bibinfo {title} {Intercomparison of pressure standards (au, pt, mo, mgo, nacl and ne) to 2.5 mbar},\ }\href {https://doi.org/10.1029/2012JB009292} {\bibfield  {journal} {\bibinfo  {journal} {J.\ Geophys.\ Res.}\ }\textbf {\bibinfo {volume} {117}},\ \bibinfo {pages} {B08210} (\bibinfo {year} {2012})}\BibitemShut {NoStop}%
\bibitem [{\citenamefont {Wu}\ and\ \citenamefont {Wentzcovitch}(2009)}]{wu2009}%
  \BibitemOpen
  \bibfield  {author} {\bibinfo {author} {\bibfnamefont {Z.}~\bibnamefont {Wu}}\ and\ \bibinfo {author} {\bibfnamefont {R.~M.}\ \bibnamefont {Wentzcovitch}},\ }\bibfield  {title} {\bibinfo {title} {Effective semiempirical ansatz for computing anharmonic free energies},\ }\href {https://doi.org/10.1103/PhysRevB.79.104304} {\bibfield  {journal} {\bibinfo  {journal} {Phys.\ Rev.\ B}\ }\textbf {\bibinfo {volume} {79}},\ \bibinfo {pages} {104304} (\bibinfo {year} {2009})}\BibitemShut {NoStop}%
\bibitem [{\citenamefont {Schubert}\ \emph {et~al.}(1975)\citenamefont {Schubert}, \citenamefont {Yuen},\ and\ \citenamefont {Turcotte}}]{schubert1975}%
  \BibitemOpen
  \bibfield  {author} {\bibinfo {author} {\bibfnamefont {G.}~\bibnamefont {Schubert}}, \bibinfo {author} {\bibfnamefont {D.~A.}\ \bibnamefont {Yuen}},\ and\ \bibinfo {author} {\bibfnamefont {D.~L.}\ \bibnamefont {Turcotte}},\ }\bibfield  {title} {\bibinfo {title} {Role of phase transitions in a dynamic mantle},\ }\href {https://doi.org/10.1111/j.1365-246X.1975.tb05888.x} {\bibfield  {journal} {\bibinfo  {journal} {Geophys.\ J.\ Int.}\ }\textbf {\bibinfo {volume} {42}},\ \bibinfo {pages} {705} (\bibinfo {year} {1975})}\BibitemShut {NoStop}%
\bibitem [{\citenamefont {Faccenda}\ and\ \citenamefont {Zilio}(2017)}]{faccenda2017}%
  \BibitemOpen
  \bibfield  {author} {\bibinfo {author} {\bibfnamefont {M.}~\bibnamefont {Faccenda}}\ and\ \bibinfo {author} {\bibfnamefont {L.~D.}\ \bibnamefont {Zilio}},\ }\bibfield  {title} {\bibinfo {title} {The role of solid--solid phase transitions in mantle convection},\ }\href {https://doi.org/10.1016/j.lithos.2016.11.007} {\bibfield  {journal} {\bibinfo  {journal} {Lithos}\ }\textbf {\bibinfo {volume} {268--271}},\ \bibinfo {pages} {198} (\bibinfo {year} {2017})}\BibitemShut {NoStop}%
\bibitem [{\citenamefont {Christensen}(1995)}]{christensen1995}%
  \BibitemOpen
  \bibfield  {author} {\bibinfo {author} {\bibfnamefont {U.}~\bibnamefont {Christensen}},\ }\bibfield  {title} {\bibinfo {title} {Effects of phase transitions on mantle convection},\ }\href {https://doi.org/10.1146/annurev.ea.23.050195.000433} {\bibfield  {journal} {\bibinfo  {journal} {Annu.\ Rev.\ Earth Planet.\ Sci.}\ }\textbf {\bibinfo {volume} {23}},\ \bibinfo {pages} {65} (\bibinfo {year} {1995})}\BibitemShut {NoStop}%
\bibitem [{\citenamefont {Christensen}(2018)}]{christensen2018}%
  \BibitemOpen
  \bibfield  {author} {\bibinfo {author} {\bibfnamefont {U.}~\bibnamefont {Christensen}},\ }\bibfield  {title} {\bibinfo {title} {Geodynamo models with a stable layer and heterogeneous heat flow at the top of the core},\ }\href {https://doi.org/10.1093/gji/ggy352} {\bibfield  {journal} {\bibinfo  {journal} {Geophys.\ J.\ Int.}\ }\textbf {\bibinfo {volume} {215}},\ \bibinfo {pages} {1338} (\bibinfo {year} {2018})}\BibitemShut {NoStop}%
\bibitem [{\citenamefont {Hoolst}\ \emph {et~al.}(2019)\citenamefont {Hoolst}, \citenamefont {Noack},\ and\ \citenamefont {Rivoldini}}]{vanhoolst2019}%
  \BibitemOpen
  \bibfield  {author} {\bibinfo {author} {\bibfnamefont {T.~V.}\ \bibnamefont {Hoolst}}, \bibinfo {author} {\bibfnamefont {L.}~\bibnamefont {Noack}},\ and\ \bibinfo {author} {\bibfnamefont {A.}~\bibnamefont {Rivoldini}},\ }\bibfield  {title} {\bibinfo {title} {Exoplanet interiors and habitability},\ }\href {https://doi.org/10.1080/23746149.2019.1630316} {\bibfield  {journal} {\bibinfo  {journal} {Adv.\ Phys.\ X}\ }\textbf {\bibinfo {volume} {4}},\ \bibinfo {pages} {1630316} (\bibinfo {year} {2019})}\BibitemShut {NoStop}%
\bibitem [{osf()}]{osf_y9qnm}%
  \BibitemOpen
  \href {https://doi.org/10.17605/OSF.IO/Y9QNM} {}\bibinfo {howpublished} {\url{https://osf.io/y9qnm} with DOI: 10.17605/OSF.IO/Y9QNM}\BibitemShut {NoStop}%
\end{thebibliography}%
\end{document}


\title{Supplemental Material for\\``Melting phase relation of seifertite and pyrite-type SiO\textsubscript{2}\\ determined by machine learning potentials''}%

\author{Doyoon Park$^{1}$}
\author{Xin Deng$^{2}$}
\email{xdeng@carnegiescience.edu}
\author{Jie Deng$^{3}$}
\email{jie.deng@princeton.edu}
\affiliation{
\textsuperscript{1}\mbox{Department of Mechanical and Aerospace Engineering, Princeton University, Princeton, New Jersey 08544, USA} \\
\textsuperscript{2}\mbox{Earth and Planets Laboratory, Carnegie Institution for Science, Washington, DC, 20015, USA} \\
\textsuperscript{3}Department of Geosciences, Princeton University, Princeton, New Jersey 08544, USA
}

\date{\today}
\maketitle

\renewcommand{\thefigure}{S\arabic{figure}}

\begin{figure*}[h]
\centering
    \includegraphics[width=0.85\textwidth]{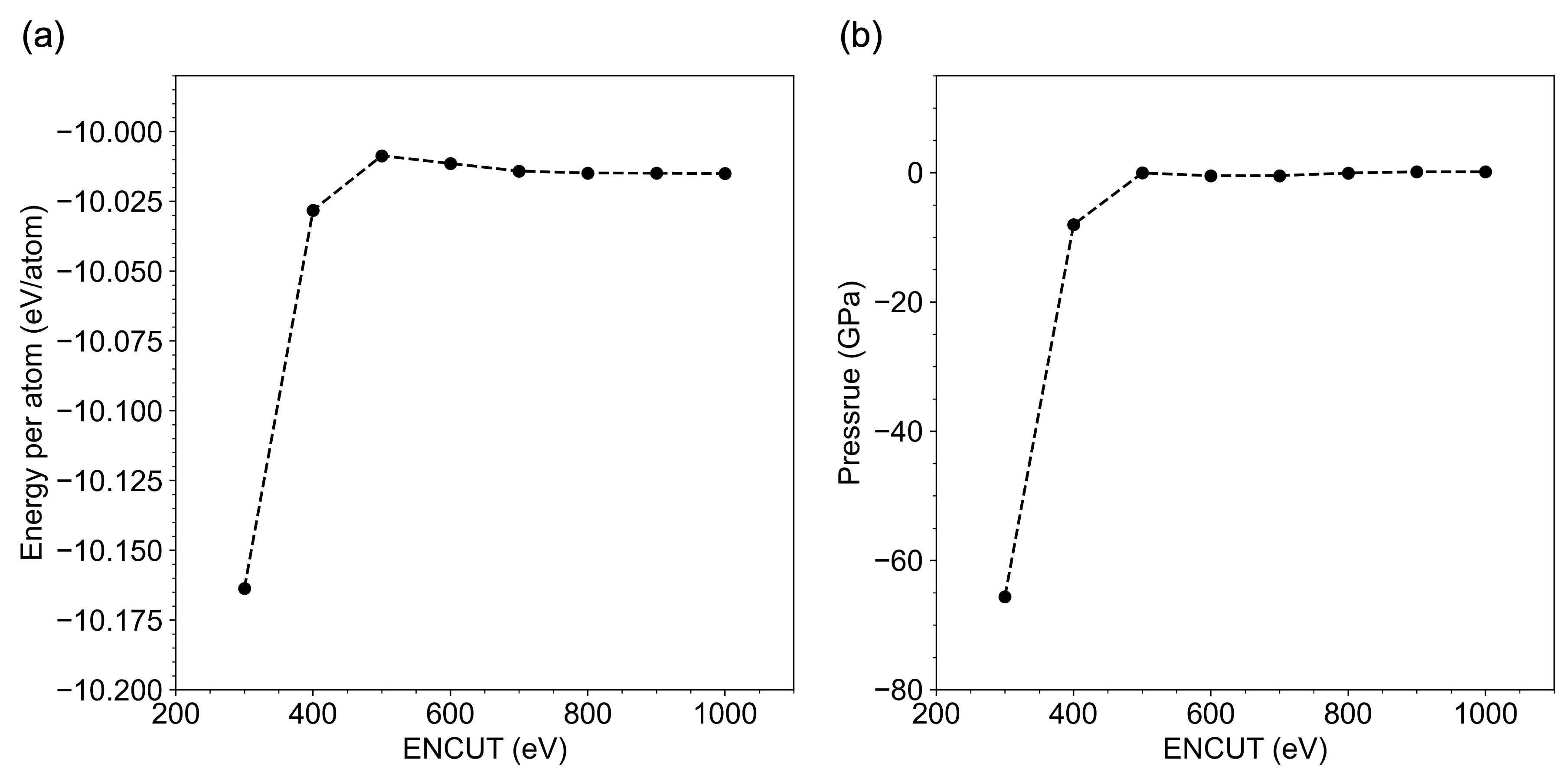}
\caption{\label{fig:s3}Convergence test of energy (a) and pressure (b) with varying energy cutoff (ENCUT) for the SiO$_2$ system containing 96 atoms. An energy cutoff of 800 eV was chosen to obtain converged results for both energy and pressure.}
\end{figure*}

\begin{figure*}[t]
\centering
    \includegraphics[width=0.90\textwidth]{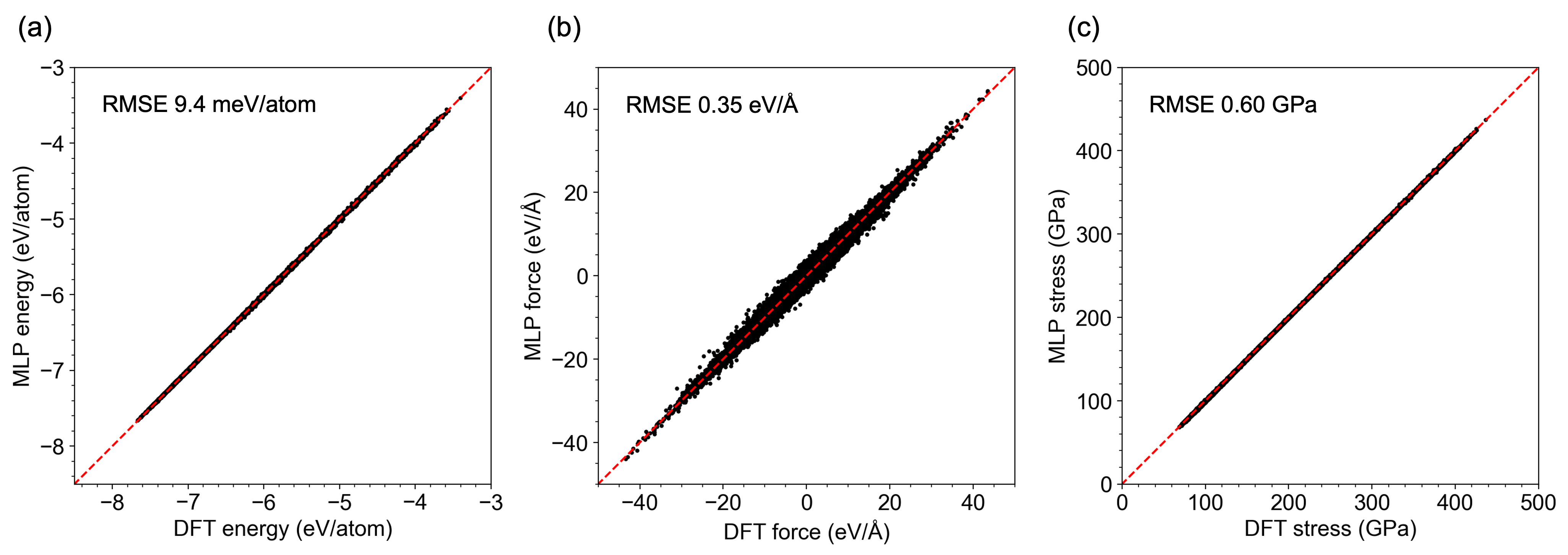}
\caption{\label{fig:s1}Comparison between MLP-PBEsol predictions and DFT calculations for energies (a), atomic forces (b), and stresses (c) using a test dataset of 10400 96-atom SiO$_2$ configurations over the temperature range 1000–10000 K and pressure range 100–400 GPa. The red dashed lines are given as guides for perfect matches.}
\end{figure*}

\begin{figure*}[b]
\centering
    \includegraphics[width=0.5\textwidth]{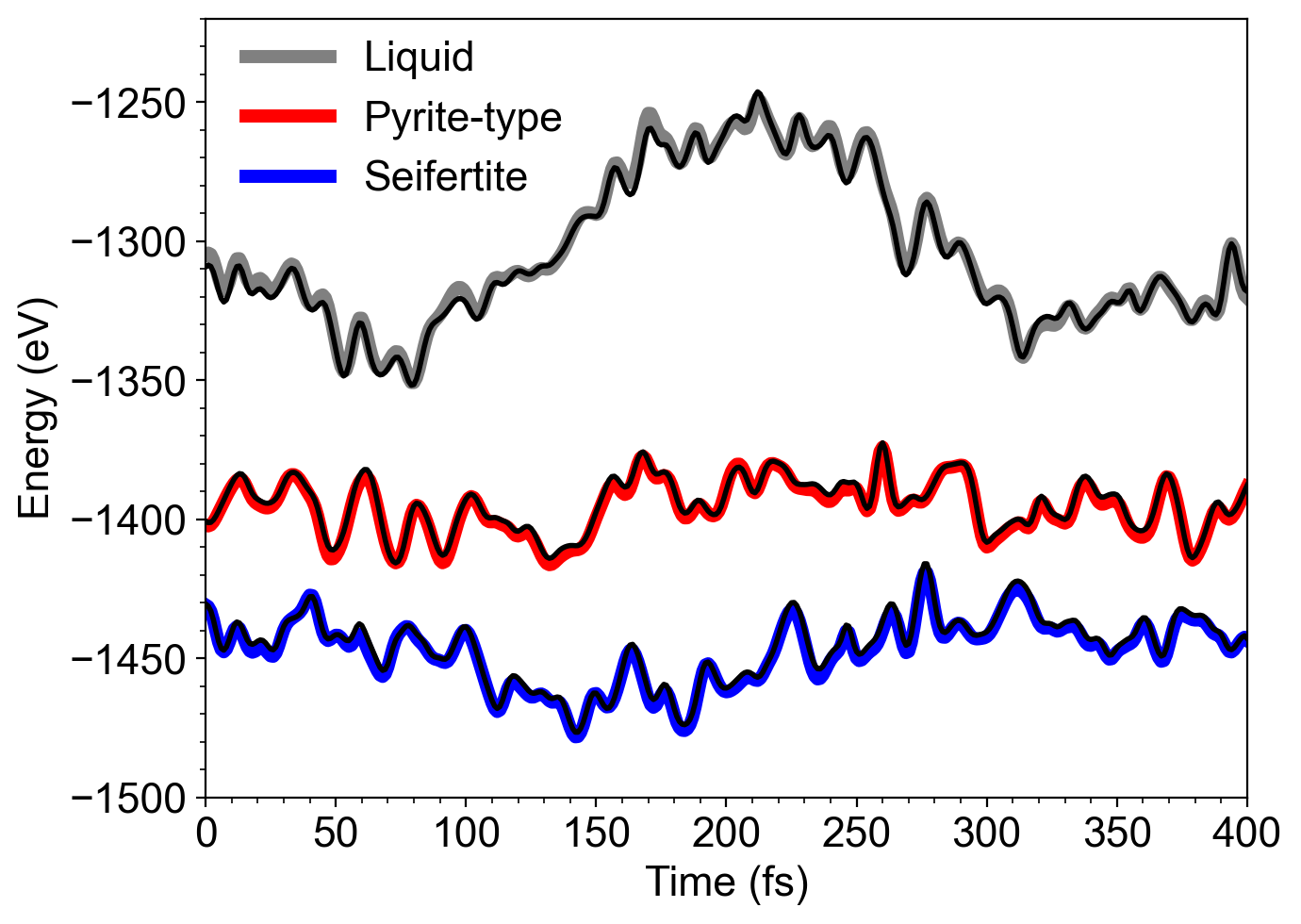}
\caption{\label{fig:s2}Comparison between MLP-PBEsol predictions (thin black lines) and DFT calculations (thick colored lines) of total energies for SiO$_2$ systems with 216 atoms at 150 GPa and 7000 K. None of the configurations in the trajectory were included in the training set. The root-mean-square errors of the MLP are 9.2, 7.9, and 9.8 meV/atom for seifertite, pyrite-type, and liquid, respectively.}
\end{figure*}

\begin{figure*}[t]
\centering
    \includegraphics[width=0.85\textwidth]{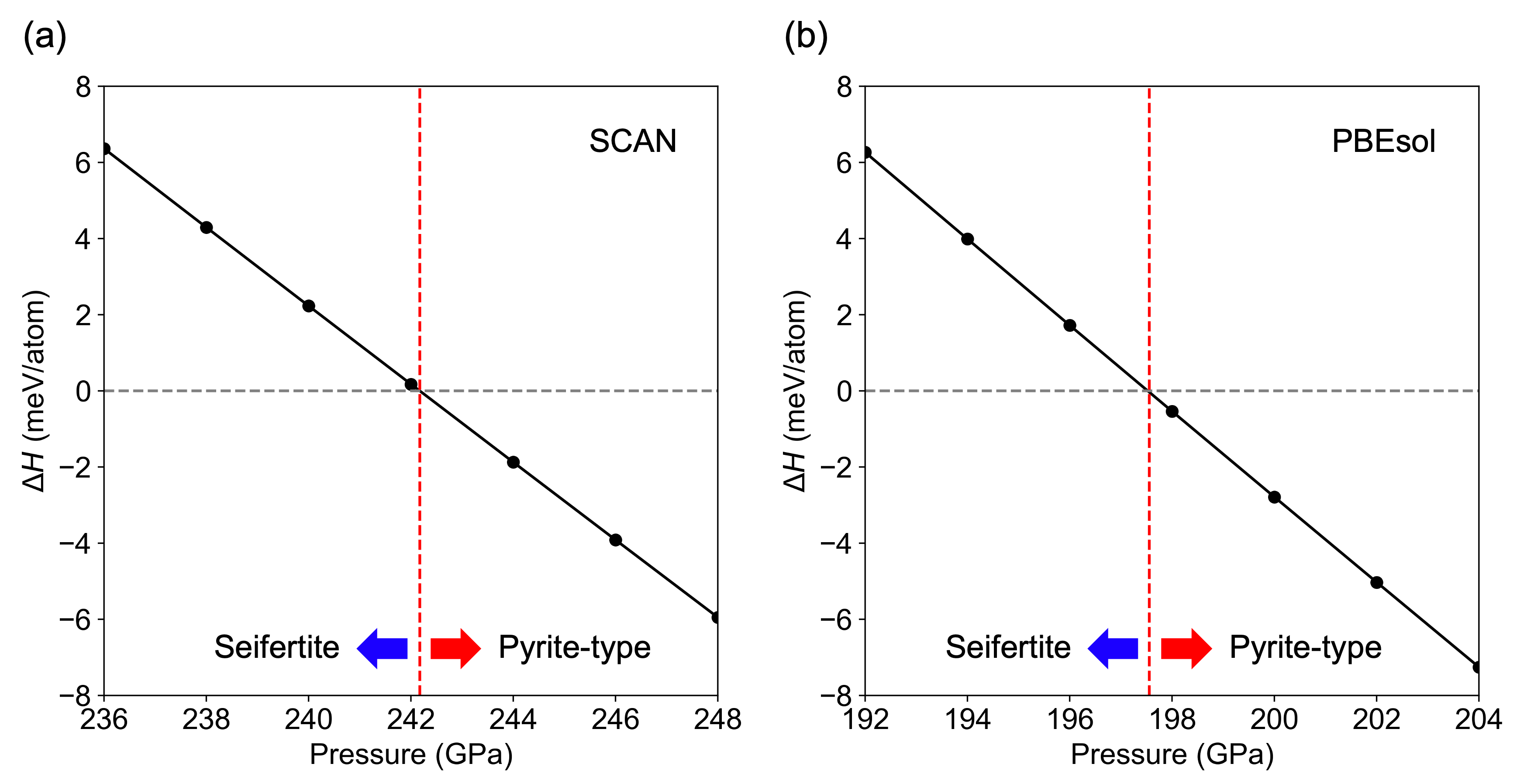}
\caption{\label{fig:s4}Enthalpy difference ($H_{\text{seifertite}}-H_{\text{pyrite}}$) between seifertite and pyrite-type SiO$_2$ as a function of pressure at 0 K, determined using the SCAN (a) and the PBEsol exchange-correlation functionals (b), respectively.}
\end{figure*}

\begin{figure*}[b]
\centering
    \includegraphics[width=0.65\textwidth]{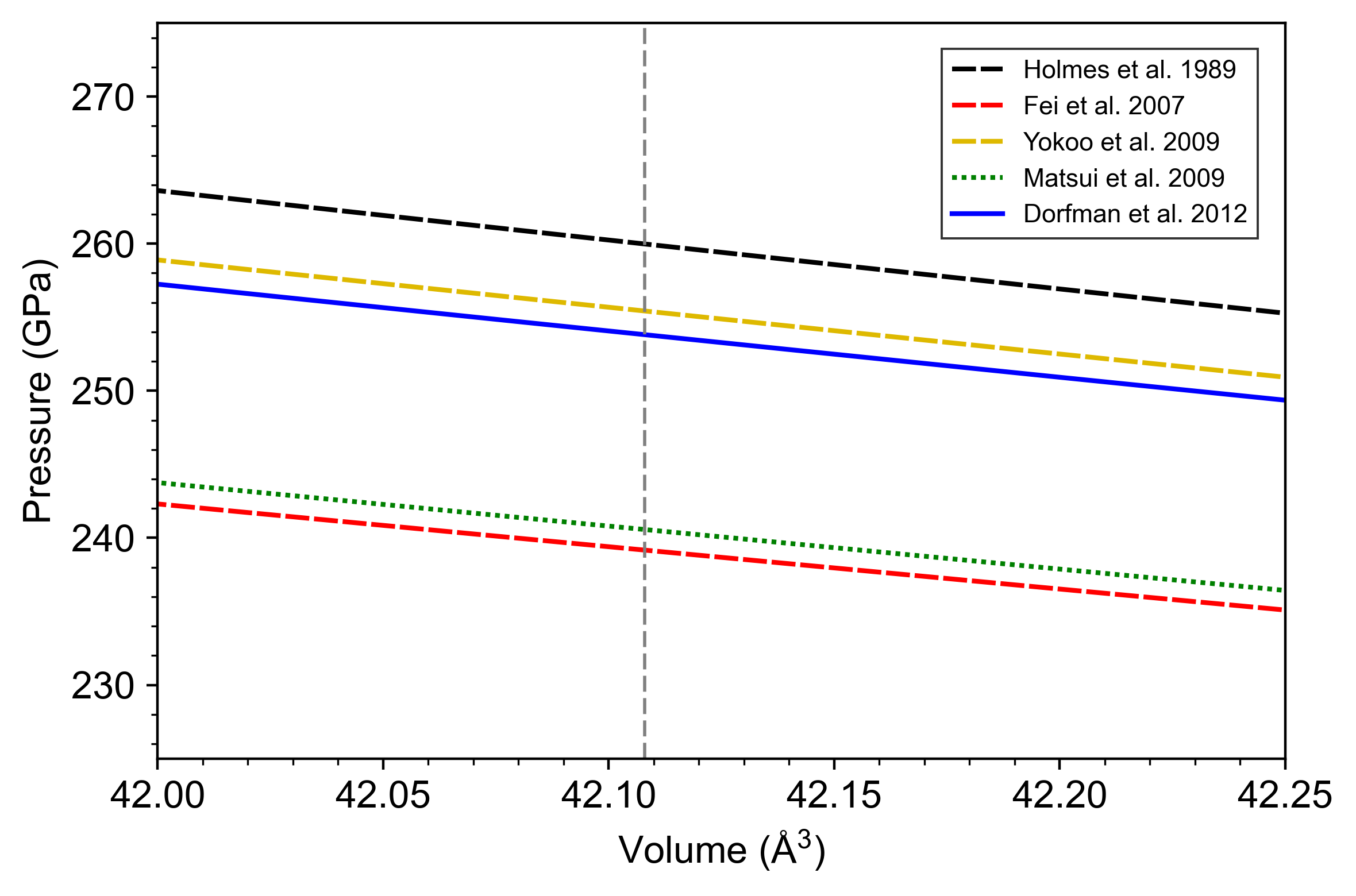}
\caption{\label{fig:s5}Comparison of equations of state (EOS) for Pt. EOSs determined from shock experiments are shown as dashed lines \cite{holmes1989, fei2007, yokoo2009}, while the EOS determined from x-ray diffraction is shown as a solid line \cite{dorfman2012}. The theoretical study is represented by a dotted line \cite{matsui2009}. The gray vertical dashed line indicates the volume at which the EOS of Holmes et al. \cite{holmes1989} reaches 260 GPa, whereas the other EOSs predict lower pressures, ranging from about 240 to 255 GPa, at the same volume.}
\end{figure*}

\clearpage
\bibliography{Supplemental}
\nocite{*}